\documentclass[10pt,twocolumn,romanappendices,journal]{IEEEtran}

\usepackage[T1]{fontenc}		% Encoding
\usepackage[utf8]{inputenc}	% Write input

\usepackage{cite}
\usepackage{graphicx}
\usepackage{amsmath,amssymb}
\usepackage{amsfonts}

\usepackage{comment}
\usepackage{epstopdf}
\usepackage{color}

\usepackage{hyperref}
\usepackage{subcaption}
\usepackage{nicefrac}
\usepackage{placeins} % \FloatBarrier
\usepackage[dvipsnames]{xcolor}

\usepackage{microtype}

\usepackage{makecell}
\usepackage{stfloats}
\usepackage{stfloats}
\usepackage{bm}

\usepackage{amssymb}
\usepackage{indentfirst}
\usepackage{extarrows}

\usepackage{cite}
\usepackage{graphicx}
\usepackage{amsthm}
\usepackage{amsmath}

\usepackage{comment}
\usepackage{epstopdf}
\usepackage{color}

\usepackage{subcaption}
\usepackage{nicefrac}
\usepackage{placeins} % \FloatBarrier
\usepackage{algorithm}
\usepackage{algorithmic}
\usepackage[dvipsnames]{xcolor}
\usepackage{diagbox}

\newtheorem{remark}{Remark}
\newtheorem{theorem}{Theorem}

\newtheorem{prop}{Proposition}

\newtheorem{example}{Example}

\usepackage{hyperref}
\hypersetup{
	colorlinks = true,
	linkcolor  = red,
	citecolor  = green!80!black,
	urlcolor   = black
}

\usepackage{url}

\definecolor{color1}{rgb}{0.00000,0.44700,0.74100}%
\colorlet{myred}{red!80!black}%
\colorlet{myblue}{color1!80!black}%
\colorlet{green}{green!80!black}%

\newenvironment{inlineproof}{\noindent\textit{Proof:}~}{\hfill $\blacksquare$ \par}
\usepackage{svg}

\begin{document}

\title{{Caching Yields up to 5× Spectral Efficiency in Multi-Beam Satellite Communications}}
\author{Hui Zhao,  Dirk Slock, and Petros Elia
	\thanks{Hui Zhao,  Dirk Slock and Petros Elia are with the Communication Systems Department, EURECOM, 06410 Sophia Antipolis, France (email: hui.zhao@eurecom.fr; dirk.slock@eurecom.fr; elia@eurecom.fr). 
	}
}

%\markboth{IEEE Wireless Communications Letters}{}
%{Shell \MakeLowercase{\textit{et al.}}: Bare Demo of IEEEtran.cls for Journals}

\maketitle

\begin{abstract}This paper examines the integration of vector coded caching (VCC) into multi-beam satellite communications (SATCOM) systems and demonstrates that even limited receiver-side caching can substantially enhance spectral efficiency. By leveraging cached content to suppress interference, VCC enables the concurrent transmission of multiple precoded signal vectors that would otherwise require separate transmission resources. This leads to a multiplicative improvement in resource utilization in SATCOM. To characterize this performance, we model the satellite-to-ground channel using Rician-shadowed fading and after incorporating practical considerations such as matched-filter precoding, channel state information (CSI) acquisition overhead as well as CSI imperfections at the transmitter, we here derive closed-form expressions for the average sum rate and spectral efficiency gain of VCC in SATCOM. Our analysis, tightly validated through numerical simulations, reveals that VCC can yield spectral efficiency gains of 300\% to 550\% over traditional multi-user MISO SATCOM with the same resources. These gains---which have nothing to do with multicasting, prefetching gains nor file popularity---highlight VCC as a pure physical-layer solution for future high-throughput SATCOM systems, significantly narrowing the performance gap between satellite and wired networks.
\end{abstract}

\begin{IEEEkeywords}Vector coded caching,  satellite communications, spectral efficiency, Rician-shadowed fading.
\end{IEEEkeywords}

%%%%%%%%%%%%%%%%%%%%%%%%%%%%%%%%%%
\section{Introduction}
%%%%%%%%%%%%%%%%%%%%%%%%%%%%%%%%%%
The rapid growth of satellite communications (SATCOM) traffic has created an urgent demand for higher spectral efficiency. To address this, \emph{multi-beam satellites} have gained significant attention~\cite{Khammassi}, employing an array of $L$ radio-frequency (RF) feeds to generate multiple beams that reuse the same frequencies. Naturally many challenges remain---like for example that of \emph{inter-beam interference}~\cite{Bhavani}---leading to a fundamental \emph{``few-feeds–many-users'' mismatch}~\cite{Cong} where simply the number of active ground terminals $K$ often far exceeds the number of available feeds $L$, thus resulting in precoding-enabled multi-beam SATCOM systems that fall short of the spectral efficiency required to support rapidly growing user demands of large content.

%The rapid growth of satellite communications (SATCOM) traffic has intensified the quest for ever more efficient spectrum-reuse strategies. In this context, \emph{multi-beam satellites} have attracted considerable attention~\cite{Khammassi}, employing an array of $L$ radio-frequency (RF) feeds to illuminate the Earth with multiple beams that reuse the same frequencies. Beam spill-over, however, gives rise to pronounced \emph{inter-beam interference}, which motivates the adoption of linear precoding techniques---such as the minimum mean square error (MMSE)~\cite{Bhavani}---at the gateway or on-board processor. Yet a persistent \emph{``few-feeds--many-users'' mismatch} remains~\cite{Cong}, as the number of concurrently active ground terminals $K$ typically far exceeds the available feeds $L$ (i.e., $K \gg L$), so that even precoding-enabled multi-beam SATCOM continues to fall short of the spectral efficiency targets required to support rapidly growing user demands of large content.

At the same time, this growth of traffic is largely due to Video on Demand (VoD), and thus, as one would expect, caching has been used as a means of alleviating this traffic. However, due to the relatively modest size of receiver-side caches, as compared to the immense libraries of content, traditional prefetching techniques result in very modest gains~\cite{Ali,9427217,7959863,8007072}.  Even the modern techniques of coded caching~\cite{Ali} that cleverly leverage receiver-side caching and coded multicast transmissions to directly handle interference,  do not provide substantial spectral efficiency gains in SATCOM networks with large antenna arrays and many beams, mainly because such coded caching solutions were intended for single-antenna settings. Additionally, even advanced \emph{multi-antenna coded caching} schemes~\cite{Shariatpanahi_CC} offer limited gains over practical multi-beam downlink systems. Consider a downlink system with \(L\) transmit antennas/feeds serving \(K\) receivers, each with a cache of size equal to a fraction \(\gamma \in [0,1]\) of the entire library, and in the presence of \(\Lambda\) distinct cache states.\footnote{This aspect of cache states is an esoteric aspect of coded caching, which though turns out to be its Achilles hill.} Most multi-antenna coded caching approaches (e.g.,~\cite{9163148,MohammadJavadTWCwithus2021,Tolli2020,Meixia_Tao_TWC2019}) achieve a Degrees-of-Freedom (DoF) performance of \(L + \Lambda \gamma\), corresponding to a caching gain \(\Lambda \gamma \ll L\) that is forced to be very small due to subpacketization (file-size) constraints~\cite{Lampiris2018_JSAC}. As a result, practical DoF performance remains close to \(L\), offering very modest improvement over cacheless multi-user (MU) multiple-input and single-output (MISO) baselines.

Vector coded caching (VCC) fundamentally transforms this scenario by enabling a multiplicative improvement in spectral efficiency compared to traditional MU-MISO \cite{Lampiris2018_JSAC,Zhao_VectorCC}. Rather than using the widely adopted XOR-based coded multicasting, VCC instead fuses multiple precoded $L \times 1$ signal vectors into a single superimposed $L \times 1$ signal vector. This superimposed vector, as well as the cache placement, are carefully designed so that the interference---resulting from overloading the system with multiple vectors in one shot---can be `cached-out' at the receiver side. Thus, VCC allows for great efficiencies compared to traditional MU-MISO systems that would require each precoded signal vector to be sent \emph{sequentially}.  This new VCC approach overcomes the constraints of conventional multi-antenna coded caching, and---by exploiting both caching and antenna resources---achieves a theoretical DoF of $L ( \Lambda \gamma + 1)$, yielding multiplicative DoF gains over downlink MU-MISO systems. 

In exploring the actual throughput gains of VCC, recent studies (cf. \cite{EURECOM+7083,Zhao2023_WSASCC,Zhao_VCC_TWC2}) have confirmed the effectiveness of VCC in urban Macro-cell and Micro-cell environments under various realistic considerations, including finite file sizes (i.e., modest values of $\Lambda\gamma$), various fading and pathloss characteristics, and various costs of acquiring channel state information (CSI). In this context, VCC demonstrates a \emph{multiplicative} improvement in spectral efficiency over independently optimized state-of-art MU-MISO terrestrial networks, validating its effectiveness in complex, high-density deployment scenarios.

{\color{black}{Despite the reported performance gains in terrestrial networks \cite{EURECOM+7083,Zhao2023_WSASCC,Zhao_VCC_TWC2}, the application of VCC in SATCOM remains largely unexplored.  It is worth noting that VCC is fundamentally different from prior SATCOM coded caching works (e.g., \cite{10731897,10347487}), where satellites collectively store the library and users retrieve files by connecting sequentially over an orbital cycle using XOR-based coded multicasting \cite{Ali,Ali_decentralized} to achieve global coverage.}}

SATCOM is not only a pertinent setting for applying VCC, but---as it turns out---also enjoys  a privileged relationship with VCC. % a privil are encouraged by obvious synergies between SATCOM and VCC. %For example, VCC thrives in the presence of many users with statistically symmetric links, which is at the core of SATCOM systems that indeed offer a near-uniformity in link conditions {\color{red}{over a terrestrial coverage with radius ranging from several tens of kilometers to 100 kilometers \cite{e26050419,Chondrogiannis,Zhao_ICC22}}}, and a coverage of a much larger number of users compared to cellular settings. 
To begin with,  SATCOM is heavily constrained by low spectral efficiency---which VCC directly improves---and its most recent advancements are closely tied to VoD \cite{Zhao_VectorCC,Zhao_VCC_TWC2}. More interestingly, we also see certain synergies.  For example, VCC thrives in the presence of many users with statistically symmetric links; two characteristics that are inherent in SATCOM systems, where link uniformity extends over terrestrial coverages with radii of several tens to one hundred kilometers \cite{e26050419,Chondrogiannis,Pan_TWC} that encompasses a far larger user population compared to cellular networks.
Furthermore, SATCOM systems endure substantially degraded CSI at the transmitter (CSIT)~\cite{Bhavani,Zhao_LMS}, which plays to the strength of VCC which migrates much of the burden of interference cancellation away from the transmitter. At the same time though, SATCOM systems are dominated by reduced received SNR values, which can be particularly problematic for VCC which operates by simultaneously serving multiple precoded signal vectors, thus leaving each vector with diminished power. 

%Our analysis carefully balances all this, to reveal the effect of VCC in SATCOM systems, taking into consideration the idiosyncratic channel conditions in SATCOM~\cite{Abdi,Singh,Quynh,Eunsun,Miao_TAES,Kshitija}. 

{\color{black}{Driven by the rapidly increasing demand for higher spectral efficiency in multi-beam SATCOM networks \cite{Khammassi, Bhavani, Cong} and the aforementioned synergies between VCC and SATCOM, \emph{our aim is to analyze the extent to which VCC enhances SATCOM spectral efficiency}, while carefully accounting for the fundamentally different characteristics of SATCOM compared to traditional cellular settings.}}  
Toward this, we model the satellite-to-ground channel as a Rician-shadowed fading channel~\cite{Abdi,Singh,Quynh,Eunsun,Miao_TAES,Kshitija}, and place our focus on the practical case of matched-filter (MF) precoding \cite{Rusek}, all-while accounting for CSI acquisition costs and the effect of imperfect CSIT. 
The primary technical contribution of this work is the \emph{parametrization of VCC delivery performance in SATCOM systems}, achieved through simple closed-form expressions that capture both the overall system spectral efficiency and the multiplicative effective gain over the cacheless MU-MISO baseline. These analytical expressions are shown to be highly accurate when validated against Monte Carlo simulations. In the end, the results reveal that under the aforementioned practical CSIT and SNR considerations, VCC offers a very sizable multiplicative spectral efficiency gain compared to optimized conventional MU-MISO, highlighting its strong potential for future high-demand satellite networks.

\emph{Notations}: For a positive integer \( n \), we use \([n]\) to denote the set \(\{1, 2, \dots, n\}\). For two sets \( \mathcal{A} \) and \( \mathcal{B} \), the notation \( \mathcal{A} \setminus \mathcal{B} \) represents the set difference between \( \mathcal{A} \) and \( \mathcal{B} \). The operator \( |\cdot| \) represents the cardinality of a set or the magnitude of a complex number, depending on the context. \( \mathbb{E}\{\cdot\} \)  denotes the expectation operator, while \( \text{Tr}\{\cdot\} \) represents the trace operator. For a matrix \( \mathbf{A} \), we use \( \mathbf{A}^T \), \( \mathbf{A}^* \), and \( \mathbf{A}^H \) to denote the transpose, element-wise complex conjugate, and conjugate transpose of \( \mathbf{A} \), respectively. The notation \( \mathbf{0}_L \) represents an \( L \times 1 \) vector of zeros, and \( \mathbf{I}_L \) denotes the \( L \times L \) identity matrix. We use \( \mathcal{CN} \) to denote the complex Gaussian distribution.

\emph{Paper Organization:} Section~\ref{Sys_model_sec} introduces the system architecture of the considered SATCOM scenario, followed by a detailed description of the VCC strategy along with the adopted precoding approach. Section~\ref{analysis_sec} focuses on the theoretical analysis of the content delivery performance under the proposed caching framework. Section~\ref{numerical_sec} presents simulation results to corroborate the theoretical findings and to demonstrate the performance gain achieved by VCC when compared to systems without caching. Finally, Section~\ref{Conclusion_sec} summarizes the main conclusions of this work. For the sake of clarity, several mathematical proofs are deferred to the Appendix.

\section{System Model}\label{Sys_model_sec}

We consider a SATCOM system in which a satellite equipped with \( L \) transmit antennas/feeds serves \( K \) cache-aided single-antenna ground users. Each user requests \emph{distinct} content files from a library \( \mathcal{F} \triangleq \{W_1, \cdots, W_N\} \) containing \( N \) equal-sized files. %\footnote{As discussed in \cite{Zhao_VectorCC} using Netflix as an example, each movie is divided into multiple equal-sized files, of size equal to several tens of megabytes each. The collection of files from top popular movies, whose delivery will be accelerated using vector coded caching, forms the library \( \mathcal{F} \). With modestly sized caches, and a cache placement that is agnostic to the file popularity of the files chosen to be one can cover a large percentage of netflix traffic.}. 
The satellite has full access to the library \( \mathcal{F} \), facilitated by a high-speed feeder link connecting the satellite to a terrestrial satellite gateway, which in turn links to the core network \cite{Zhao_ICC22}.  During an off-peak period (for example, once a month), each user can store (and occasionally update) a fraction \( \gamma \in [0, 1] \) of the library content, where $\gamma$ is the normalized cache size compared to the library size\footnote{In principle, one would consider the part of the library corresponding to a sizable fraction of the overall VoD traffic---for example 90\% of traffic (cf. \cite[Example 1]{Zhao_VectorCC}).}. During all the intermediate transmission phases (peak times), the satellite must deliver individual content to its different users, where each user gets their own content, as is typical in VoD. 

{\color{black}{We adopt a static (quasi-static) model for the satellite–terrestrial channel, where the channel coefficients remain constant within a transmission block and may vary independently across different blocks.}} To justify this assumption, we consider that land terminals are located within a coverage radius of 100~km, with a circular LEO orbit at an altitude of 600~km above the coverage center. Satellites are spaced 75~km apart, corresponding to an inter-satellite central angle of $\theta \approx 0.674^\circ$. A handover is triggered once the angle between the satellite and the line connecting the Earth's center to the coverage center exceeds $\theta/2$. This corresponds to a change in the satellite's zenith angle from about $0.591^\circ$ to $1.208^\circ$ as observed by land terminals located at the opposite edge of the coverage area, indicating an extremely small variation. With precise beam alignment to the serving satellite, the relative geometry between the serving satellite and the covered terminals remains nearly unchanged. Under such conditions and  Doppler shift compensation, the LEO-to-ground channel can be regarded as being a static fading channel, similar to that of GEO satellite--terrestrial links. %{\color{blue}{This static channel assumption is further supported by the numerical results in Fig.~\ref{Gain_Dyna_fig}.

Under the static channel model, we consider the Rician-shadowed fading channel \cite{Abdi}, a common model for satellite-to-ground communication that accounts for both line-of-sight (LOS) and non-line-of-sight (NLOS) components.  This model is defined by three key parameters: the power of the scattering component, denoted by \( 2 \beta \); the power of the LOS component, represented by \( \Omega \); and the parameter \( m  \) reflecting the extent of LOS obstruction caused by environmental elements like buildings, trees, or terrain. Here, \( m = 0 \) represents a completely obstructed LOS, while \( m \to \infty \) indicates no obstruction. The Rician-shadowed fading model can accommodate different types of orbits and frequency bands (e.g., Ku-band and Ka-band) \cite{ Singh,Quynh,Eunsun,Miao_TAES,Kshitija}. 
%It should be thus noted that our performance analysis covers both LEO and GEO systems, but that indeed our interest rests mainly on the LEO case which is certainly closely associated to VoD delivery.
The channel vector between the satellite and the \( k \)-th user ($k \in [K]$) is represented by \( \mathbf{h}_k \in \mathbb{C}^{L \times 1} \),  which can be expressed as
\begin{align}
    \mathbf{h}_k  = Z_k \mathbf{t}_k  +  \mathbf{h}_k',
\end{align}
where $\mathbf{h}'_{k}\sim\mathcal{CN}(\mathbf{0}_{L},\,2\beta\,\mathbf{I}_{L})$ denotes the scattered component; 
$\mathbf{t}_{k} \triangleq [\exp(\jmath \theta_{k}^{(1)}), \cdots, \exp(\jmath \theta_{k}^{(L)})]^T$ collects the LOS phasor from the transmit antennas to user $k$; and the real-valued scalar
$Z_{k}$---following an Nakagami-$m$ distribution with shape parameter $m$ and scale parameter $\Omega$---models the LOS amplitude fluctuation. 
In the considered LEO-to-ground scenario with slowly moving users, the phase of the LOS 
component from each transmit antenna to the ground user would, in principle, be deterministic 
and highly correlated, as it is set by the array geometry and the satellite--user geometry. 
Nevertheless, there are several reasons to model these LOS phases differently. First, relative 
satellite motion, oscillator drifts, and hardware-induced phase jitter introduce effective random 
phase offsets across antennas, which decorrelate LOS contributions over time. Second, deliberate 
phase dithering is sometimes employed in practice to randomize the LOS component and reduce 
systematic interference patterns. Accordingly, for analytical tractability we adopt a block-fading 
abstraction: within a channel coherence block the LOS phases are fixed, whereas across different channel 
coherence blocks they are modeled as i.i.d.\ and uniformly distributed over $[0,2\pi)$.

In a given transmission round, let \( \mathbf{x} \in \mathbb{C}^{L \times 1} \), with $\mathbb{E}\{ ||{\bf x}||^2 \} = P_t$, denote the transmit signal from the satellite that conveys the requested messages for \emph{multiple users}.
The received signal at user \( k \) is of the form
\begin{align}
y_k = \mathbf{h}_k^T \mathbf{x} + z_k,
\end{align}
where \( z_k \sim \mathcal{CN}(0, 1) \) represents the additive white Gaussian noise (AWGN). Hence, $P_t$ here accounts for the effects of the actual transmit power, the large-scale pathloss, antenna gains, and the actual AWGN power.
%\begin{remark}
%It is noted here that, in a small abuse of notation, $P_t$ is used as the term that incorporates the effects of the transmit power, but also of the large-scale pathloss, antenna gain, and AWGN power.
%\end{remark}

{\color{black}{Next,}} we demonstrate how VCC enables the efficient use of each user’s cached content by detailing the design of the transmit signal vector $\mathbf{x}$, and showing how users exploit their local cache to decode their intended data while mitigating interference from others.

\subsection{Vector Coded Caching}\label{VCC_descrip}

\begin{algorithm}[t]
\caption{Vector Coded Caching (VCC) Scheme}
\label{alg:VCC}
\begin{algorithmic}[1]
{\color{black}{
\REQUIRE Library $\mathcal{F}=\{W_1,\dots,W_N\}$, total number of users $K$, cache size $\gamma$, number of cache states $\Lambda$, multiplexing gain $Q$, requested file indices $\{{d_{k}}\}_{k \in [K] }$
\ENSURE All users recover their requested files 

\vspace{0.5em}
\STATE \textbf{Cache Placement Phase:}
\FOR{each file $W_n$, $n\in[N]$}
    \STATE Split $W_n$ into $\binom{\Lambda}{\Lambda\gamma}$ non-overlapping and equal-sized subfiles $\{W_n^{\mathcal T}:\mathcal T\subseteq[\Lambda],|\mathcal T|=\Lambda\gamma\}$
\ENDFOR
\STATE Partition users into $\Lambda$ groups, each with $B=K/\Lambda$ users and a dedicated cache state
\FOR{each cache state $g\in[\Lambda]$}
    \STATE Users in group $g$ cache the identical content
    \[
     \mathcal{Z}_{g} = \{ W_n^\mathcal{T} : \ \mathcal{T}  \ni g, \forall n \in [N] \}
    \]
\ENDFOR

\vspace{1.5em}
\STATE \textbf{Content Delivery Phase:}
\FOR{each subset $\Psi\subseteq[\Lambda]$ with $|\Psi|=\Lambda\gamma+1$}
    \FOR{each transmission round $r=1,\dots,B/Q$}
        \STATE Select $Q$ users from each cache state $\psi\in\Psi$
        \STATE Form symbols $s_{\psi,b}$ from subfiles $W_{d_{\psi,b}}^{\Psi\setminus\{\psi\}}$
        \STATE Construct the transmit signal
        $
        \mathbf x$ in \eqref{x_VCC}
          \FOR{each received signal \eqref{received_signal_VCC}}
             \STATE Cancel inter-group interference using cached content and composite CSI
             \STATE Decode desired symbol $s_{\psi,b}$ by handling intra-group interference via precoding
          \ENDFOR     
    \ENDFOR
\ENDFOR
}}
\end{algorithmic}
\end{algorithm}

{\color{black}{This subsection outlines the core design of VCC.}}
Similar to conventional cache-aided communications, VCC operates in two phases, namely the \textit{cache placement phase} and the \textit{content delivery phase}, {\color{black}{as summarized in \textbf{Algorithm~\ref{alg:VCC}}}}. We describe each phase in detail below.

\subsubsection{Cache Placement Phase} 
During the placement phase, each file \( W_n \) from the library \( \mathcal{F} = \{ W_1, W_2, \dots, W_N \} \) is divided into $\Lambda \choose \Lambda \gamma$ non-overlapping and equal-sized subfiles, each labelled by some $\Lambda \gamma$-tuple $\mathcal{T}$ with $ \mathcal{T} \subseteq [\Lambda]$. Users are partitioned into \( \Lambda \) user groups, where each group \( g \in [\Lambda] \) contains \( B = \frac{K}{\Lambda} \) users, %\footnote{For clarity, we restrict our discussion to the case where \( B \) is a positive integer; however, we note that VCC is also applicable when \( K \) is not an exact multiple of \( \Lambda \) \cite{Lampiris2018_JSAC,Zhao_VectorCC}.}
all caching the same subfiles. Specifically, for users in group \( g \), the cached content is of the form
 $
   \mathcal{Z}_{g} = \{ W_n^\mathcal{T} : \mathcal{T} \subseteq [\Lambda], |\mathcal{T} | = \Lambda \gamma, \mathcal{T}  \ni g, \forall n \in [N] \}.
$
It is straightforward to verify that this placement scheme adheres to the storage capacity \( \gamma \) at each user\footnote{We do not account for the cost of placement, which only occurs rarely (once a month perhaps), incrementally, and certainly during off-peak hours.  This cost does not match the constant cost of content delivery, especially during the crucial peak hours.}.
We refer to users storing the same content as being in the same \emph{cache state}.  
This setup ensures that users with distinct cache states store some overlapping content, enabling interference management during the delivery phase. 

\subsubsection{Content Delivery Phase} 
In the delivery phase, there are $\Lambda \choose \Lambda \gamma +1$ transmission stages. Each stage comprises a unique selection of cache states \( \Psi \subseteq [\Lambda] \) with \( |\Psi| \triangleq G = \Lambda \gamma + 1 \), with \( Q \) users sharing the same cache state selected for service in each transmission round. Here, $Q$ can be interpreted as the multiplexing gain provided by transmit antennas. Thus, there are \( B/Q \) transmission rounds in each stage. We use ${\rm U}_{\psi,b}$ to denote the $b$-th \emph{active} user in the user-group $\psi$ for any $b \in [Q]$ and $\psi \in [\Lambda]$, and
$d_{\psi,b} \in [N]$ denotes the file index requested by user ${\rm U}_{\psi,b}$. The signal vector \( \mathbf{x} \in \mathbb{C}^{L \times 1} \) in this transmission round is formulated as
   \begin{equation}\label{x_VCC}
   \mathbf{x}  = \alpha \sum\nolimits_{\psi \in \Psi} \sum\nolimits_{b \in [Q]} \mathbf{v}_{\psi,b} {s}_{\psi,b}  =  \alpha \sum\nolimits_{\psi \in \Psi} \mathbf{V}_{\psi}  \mathbf{s}_{\psi},
   \end{equation}
where \( {s}_{\psi,b} \in \mathbb{C}\), generated from the subfile $W_{d_{\psi,b}}^{\Psi \setminus \{\psi\} }$, represents the signal symbol intended for user ${\rm U}_{\psi,b}$, and \( \mathbf{v}_{\psi,b} \in \mathbb{C}^{L \times 1} \) is the precoding vector for ${s}_{\psi,b}$. In the above, ${\bf s}_\psi \triangleq [{s}_{\psi,1},\cdots,{s}_{\psi,Q}]^T \in \mathbb{C}^{Q \times 1}$ and ${\bf V}_\psi \triangleq [ \mathbf{v}_{\psi,1}, \cdots,  \mathbf{v}_{\psi,Q} ] \in \mathbb{C}^{L \times Q}$ denote the signal vector and the corresponding precoding matrix for the $Q$ cache-sharing users. Moreover, $\alpha$ is responsible for normalizing the average power of ${\bf x}$ into $P_t$.

The received signal at user ${\rm U}_{\psi,b}$ is expressed as \eqref{received_signal_VCC}, shown at the top of this page.
\begin{figure*}
   \begin{equation}\label{received_signal_VCC}
   {y}_{\psi,b} = \mathbf{h}_{\psi,b}^T \mathbf{x} + {z}_{\psi,b} = \alpha \mathbf{h}_{\psi,b}^T \mathbf{v}_{\psi,b}  {s}_{\psi,b} +  \underbrace{\alpha \mathbf{h}_{\psi,b}^T \sum\nolimits_{b' \neq b, b' \in [Q]} \mathbf{v}_{\psi,b'}  {s}_{\psi,b'}}_{\text{intra-group interference}} + \underbrace{\alpha  \mathbf{h}_{\psi,b}^T \sum\nolimits_{\phi \neq \psi, \phi \in \Psi} \mathbf{V}_{\phi} \mathbf{s}_{\phi}}_{\text{inter-group interference}} + {z}_{\psi,b}
   \end{equation}
 \rule{18cm}{0.01cm} 
\end{figure*}
According to the cache placement design,  user ${\rm U}_{\psi,b}$ has cached the subfiles \( \{W_{n}^{\Psi \setminus \{\phi\} } :  \  \phi \in \Psi \setminus \{\psi\}, \forall n \in [N]\} \), which will be used to cancel inter-group interference from other groups \( \phi \in \Psi \setminus \{\psi\} \).
Furthermore, the intra-group interference among users within the same group can be managed by designing the precoder \( \mathbf{V}_{\psi} \) appropriately (cf. Section~\ref{MF_subsec}). This setup provides a theoretical multiplexing gain represented by the DoF \( G Q  \), achieving a multiplicative spectral efficiency boost over traditional cacheless MU-MISO systems. After completing \( \binom{\Lambda}{\Lambda \gamma + 1} \) stages, all users receive their requested files \emph{in full}. We refer to \cite{Lampiris2018_JSAC,Zhao_VectorCC} for more details.

{\color{black}{
\begin{remark}
At the receiver side, VCC requires additional processing to suppress interference that cannot be handled by conventional precoding alone. Unlike decoding-based physical-layer (PHY) techniques such as non-orthogonal multiple access (NOMA), this cache-aided interference suppression does not rely on decoding the interfering streams. Instead, the receiver exploits its cached content and the composite CSI to locally regenerate the known interfering components and subtract them \emph{jointly} from the received signal. This operation involves only linear processing and constitutes the primary additional PHY-layer complexity introduced by VCC. The associated composite CSI cost will be accounted for in the CSI overhead parameter $\xi_{G,Q}$ (cf.  \eqref{Sum_rate_def}) when evaluating the performance gains over the cacheless baseline.
\end{remark}
}}

\begin{remark}
  Unlike content delivery networks (CDNs), which pre-position entire files based on demand prediction, VCC does not bypass transmission by guessing user preferences or pre-loading complete video content. Instead, it uses receiver-side storage to enable PHY multiplexing gains: specifically, to support the decoding of densely packed unicast streams delivered simultaneously through structured precoding. The local storage is thus employed as an enabler of advanced signal separation---allowing multiple uniquely addressed streams to be resolved from a shared transmission. The speedup in spectral efficiency will not arise from pre-downloading content, but rather from a fundamental improvement in the precoding structure. %Storage is used to support VCC---not to cheat the metric by caching. 
\end{remark}

\begin{comment}
    {\color{red}{
\begin{remark}
    It is worth noting that several works have investigated coded caching in SATCOM (e.g., \cite{10731897,10347487}). In these studies, each satellite stores a portion of the library, and as the satellites move along their orbit, users sequentially connect to different satellites. After a full orbital rotation, users around the globe can collectively retrieve their requested files. Such approaches typically rely on the XOR-based coded multicasting in \cite{Ali} or its decentralized variant \cite{Ali_decentralized}, applied across all satellites on the orbit to realize \emph{global coverage}. This multi-server architecture is fundamentally different from the VCC framework considered in this work.
\end{remark}
}}

\end{comment}

{\color{black}{Next, we describe the precoder for VCC in SATCOM.}}

\subsection{Matched-Filter (MF) Precoding}\label{MF_subsec}

We consider MF precoding due to its low complexity and favorable performance in the low-SNR regime typical of SATCOM, where it approaches the spectral efficiency of the more complex MMSE precoding~\cite{Rusek,Zhao_MF_SPAWC,Zhao_LMS}. %see also \cite{SPACE_Web,Zhao_ICC22,Zhao_LMS}.

As seen (cf.~\eqref{received_signal_VCC}), each user must have knowledge of the global CSI in order to cancel inter-group interference. The CSI dissemination process follows the standard {\color{black}{FDD}} uplink–downlink training framework (cf. \cite{Bhavani,caire2010multiuser}) and consists of three phases. In the first phase, the satellite transmits pilot symbols to the served users. In the second, each user feeds back its locally estimated CSI to the satellite gateway, typically via codebook-based quantization that conveys only a discrete index; this limited feedback inevitably leads to quantization errors at the gateway, resulting in imperfect CSIT. Finally, in the third phase, the gateway broadcasts the aggregated global CSI to all $GQ$ users served simultaneously.

Let \( \mathbf{\hat{h}}_{\psi,b} \) represent the estimated channel vector for user ${\rm U}_{\psi,b}$, obtained via the maximum likelihood (ML) estimator. The estimated channel vector can be expressed as \cite{Arti_NCC}
$
\mathbf{\hat{h}}_{\psi,b} = \mathbf{h}_{\psi,b} + \tilde{\mathbf{h}}_{\psi,b},
$
where \( \tilde{\mathbf{h}}_{\psi,b} \in \mathbb{C}^{L \times 1} \) is the estimation error. Each element of \( \tilde{\mathbf{h}}_k \) follows an i.i.d. complex Gaussian distribution with zero mean and variance \( \sigma_e^2 \). Furthermore, \( \mathbf{h}_{\psi,b} \) and \(  \tilde{\mathbf{h}}_{\psi,b} \) are  independent.

Under MF precoding, the transmit signal \( \mathbf{x} \) from \eqref{x_VCC} becomes
\begin{equation}\label{transmit_signal_MF}
\mathbf{x} =  \alpha  \sum\nolimits_{\psi \in \Psi}  \mathbf{\hat H}_{\psi}^H  \mathbf{s}_{\psi},
\end{equation}
where \( \hat {\bf H} \triangleq [ \hat{\bf h}_{\psi,1}, \cdots, \hat{\bf h}_{\psi,Q} ]^T \in \mathbb{C}^{Q \times L} \) represents the estimated channel matrix corresponding to the channel from the satellite to the selected \( Q \) users in the user-group \( \psi \), who---as we recall---share the same cached content.

{\color{black}{
\begin{remark}
The computational complexity of $ \mathbf{\hat H}_{\psi}^H$ is
$\mathcal{O}(QL)$, since it only involves a conjugate transpose operation on $\mathbf{\hat H}_{\psi}$.
In contrast, the zero-forcing (ZF) precoder
$\hat{\mathbf H}_\psi^{\mathrm H}(\hat{\mathbf H}_\psi\hat{\mathbf H}^{\mathrm H}_\psi)^{-1}$
requires the formation of a $Q\times Q$ Gram matrix, a matrix inversion, and an additional
matrix multiplication, resulting in an overall computational complexity of
$\mathcal{O}(QL+Q^{2}L+Q^{3}+LQ^{2})$.
The MMSE precoder further introduces a diagonal loading term, leading to a slightly higher computational complexity than that of ZF.
Therefore, ZF/MMSE precoding is significantly more computationally demanding than MF precoding.
\end{remark}
}}

After removing inter-group interference via locally cached content and perfect downlink channel training, the received signal at user \( {\rm U}_{\psi,b} \) who aims to decode its intended symbol \( s_{\psi,b} \), is given by
\begin{equation}
y_{\psi,b}' = \alpha  \mathbf{h}_{\psi,b}^T \hat {\bf h}_{\psi,b}^* s_{\psi,b} + \alpha  \sum_{b' \in [Q] \setminus \{b\} } \mathbf{h}_{\psi,b}^T  \hat{\bf h}_{\psi,b'}^* s_{\psi,b'} + z_{\psi,b}. \notag
\end{equation}
With common Gaussian signaling, the signal-to-interference-plus-noise ratio (SINR) for decoding \( s_{\psi,b} \) at user \( {\rm U}_{\psi,b} \) takes the form
\begin{align}
    \text{SINR}_{\psi,b} = \frac{ \alpha^2  | \mathbf{h}_{\psi,b}^T \hat {\bf h}_{\psi,b}^*|^2 }{ 1 + \alpha^2  \sum_{b' \in [Q] \setminus \{b\} }  |\mathbf{h}_{\psi,b}^T  \hat{\bf h}_{\psi,b'}^*|^2 }.
\end{align}
The effective rate for \( {\rm U}_{\psi,b} \) can be expressed as
\begin{align}
    R_{\psi,b} = \xi_{G,Q} \log_2 \left(  1 + \text{SINR}_{\psi,b} \right),
\end{align}
where the term \(\xi_{G,Q} \triangleq 1 - \frac{GQ \Theta}{T} \) accounts for the CSI acquisition overhead,  \(T\) denotes the channel coherence block length (in symbols),  and \( \Theta \) is the total pilot length per user and per block.
The effective sum rate for simultaneously serving \( GQ \) users in VCC is
\begin{align}\label{Sum_rate_def}
    R_{\rm sum}(G,Q) = \xi_{G,Q} \sum_{\psi \in \Psi} \sum_{b \in [Q]} \log_2 \left(  1 + \text{SINR}_{\psi,b} \right).
\end{align}
We note that setting \( G=1 \) corresponds to the cacheless (traditional MU-MISO) counterpart (cf. \eqref{x_VCC} and \eqref{received_signal_VCC}). To quantify the performance gain over conventional cacheless MU-MISO systems, we define the \emph{effective gain} as\footnote{This paper examines the role of precoding in cache-aided MU-MISO systems. The case \(Q = Q' = 1\), in which both VCC and the cacheless system operate without precoding, is not considered. In practice, a spot beam’s spatial resolution limits how many users can share the same time–frequency resources; typically only a few ($\le10$) can be multiplexed simultaneously, so $Q_{\max}, Q_{\max}' \le 10$, even for $L \gg 10$.}
\begin{align}\label{Gain_def}
    \mathcal{G} = \frac{ \max_{Q \in \{2,\cdots,Q_{\max}\} }  \mathbb{E}\{ R_{\rm sum}(G,Q) \}   }{  \max_{Q' \in \{2,\cdots,Q_{\max}'\} }  \mathbb{E}\{ R_{\rm sum}(1,Q') \} },
\end{align}
where the expectation is taken over the channel realizations across the statistically symmetric users.
This allows both cache-aided and cacheless schemes to achieve their \emph{optimized tradeoff} between multiplexing and beamforming gains. The effective gain \( \mathcal{G} \) thus captures the multiplicative boost in spectral efficiency over optimized conventional MU-MISO systems at finite SNRs. As one might expect, achieving the full theoretical gain of \( G \) corresponds to idealized high SNR conditions \cite{Lampiris2018_JSAC}.

\begin{remark} \label{rem:NoCheat}
    As indicated in \eqref{Gain_def},  this work does not concern it self with the local caching gain\footnote{We remind the reader (cf.~\cite{Ali}) that in practice, such gains are essentially trivial since the storage capacity of end-user devices \cite{Ali} is very small compared to the entire content library.}, which refers to the reduction in delivery load achieved by storing parts of the requested content at the user side in advance. We repeat that this means that the gains that we record, are certainly not an outcome of the fact that a fraction of the data need not be communicated because it is cached. Instead, the gains here are from comparing the delivery rates of the data that is in fact communicated during delivery. %We also clarify that VCC does not involve caching at the satellite \cite{Quynh}. In contrast, VCC provides a fundamentally different approach which enhances the spectral efficiency of PHY communications toward end users with caches.
\end{remark}

\section{Main Results}\label{analysis_sec}
In this section, we derive closed-form expressions for the average sum rate and the effective gain of VCC in SATCOM systems under imperfect CSIT, with the wireless channel modeled using Rician-shadowed fading. For analytical tractability, we assume that all users experience identical channel parameters (i.e., the same $m$, $\beta$, and $\Omega$), which preserves the generality of the insights when focusing on average system performance \cite{Eunsun,Chondrogiannis,Singh}.
%This assumption holds in scenarios where the coverage area spans several to tens of kilometers in radius, and the users, requesting the same type of service (e.g., video-on-demand) from a GEO satellite, are uniformly distributed across the region.
%However, it is important to note that the analytical method developed in this section can be readily extended to more general cases where each user has distinct channel statistics.

We first have the following result for the statistics of the estimated channel $\hat {\bf h}_{\psi,b}$.
\begin{prop}\label{Rician_shadow_prop}
    In the ML estimation of the channel vector \( \mathbf{h}_{\psi,b} \) over Rician-shadowed fading channels, each element of \( \hat{\mathbf{h}}_{\psi,b} \) shares the same statistical parameters \( m \) and \( \Omega \) as the corresponding element of \( \mathbf{h}_{\psi,b} \), differing only in the average scattering component power, which becomes \( 2\beta + \sigma_e^2 \).
\end{prop}
\begin{inlineproof}
   Let  $h_{\psi,b}^{(\ell)}$ denote the $\ell$-th element of ${\bf h}_{\psi,b}$.
    For the channel gain $h_{\psi,b}^{(\ell)}$ over Rician-shadowed fading channels, we can write it as (cf. \cite{Abdi})
    $ h_{\psi,b}^{(\ell)} = h_{\rm LOS} + h_{\rm NLOS}$
    where $h_{\rm LOS}$ represents the random fluctuation of the LOS component. In the above, $ h_{\rm NLOS}$ models the NLOS component, which follows a complex Gaussian distribution with zero-mean and variance $2 \beta$.
    For the $\ell$-th element of $\hat {\bf h}_{\psi,b}$, denoted by $\hat h_{\psi,b}^{(\ell)}$, we can write it as \cite{Arti_NCC}
    $
       \hat h_{\psi,b}^{(\ell)} = h_{\psi,b}^{(\ell)} + \tilde h_{\psi,b}^{(\ell)} =  h_{\rm LOS} + h_{\rm NLOS} + \tilde h_{\psi,b}^{(\ell)},
    $
    where $\tilde h_{\psi,b}^{(\ell)} \sim \mathcal{CN}(0, \sigma_e^2)$ denotes the ML estimation error.
    Note that $\tilde h_{\rm NLOS} \triangleq h_{\rm NLOS} + \tilde h_{\psi,b}^{(\ell)}$ follows a complex Gaussian distribution with zero-mean and variance $2 \beta + \sigma_e^2$. Thus, $\hat h_{\psi,b}^{(\ell)} =  h_{\rm LOS} + \tilde h_{\rm NLOS}$ forms a new Rician-shadowed channel gain as described in Proposition~\ref{Rician_shadow_prop}.
\end{inlineproof}

Next, we have the closed-form expression for $\alpha^2$ under MF precoding.
\begin{prop}\label{alpha_prop}
    The squared power control factor $\alpha^2$ under MF precoding with imperfect CSIT is of the form
    \begin{align}
        \alpha^2 =& \frac{P_t}{GQL (2 \beta + \sigma_e^2 + \Omega)}.
    \end{align}
    Moreover, by setting $G=1$ and $Q = Q'$ in $\alpha^2$, we have the squared power control factor $\alpha_0^2$ in the cacheless MU-MISO counterpart.
\end{prop}

\begin{inlineproof}
The proof is relegated to Appendix~\ref{alpha_prop_proof}.
\end{inlineproof}

We also have the following. 
\begin{prop}\label{Xi_prop}
For \begin{align} \label{Xi_def_eq}
    \Xi_1 \triangleq \mathbb{E}\{ ||{\bf h}_{\psi,b}||^4 \}, \ \ 
    \Xi_2 \triangleq  \mathbb{E} \{ |\mathbf{h}_{\psi,b}^T  \hat{\bf h}_{\psi,b'}^*|^2 \},
\end{align} then
    \begin{align}\label{X11_prop_eq}
        &\Xi_1 = L \left[\left( 4 \beta^2 + 4 \beta \Omega +\frac{\Omega^2}{m} \right) + \big( 2\beta + \Omega \big)^2  \right] \notag\\
        &\hspace{1cm} + L(L-1) \left[ \Big(1+\frac{1}{m} \Big) \Omega^2 + 4 \beta \Omega  + 4 \beta^2\right] \\
        &\Xi_2 = \! L (2\beta + \Omega) (2\beta + \sigma_e^2 + \Omega).
         \label{Xi2_prop_eq}
    \end{align}
\end{prop}
\begin{inlineproof}
The proof is relegated to Appendix~\ref{Xi_prop_proof}.
\end{inlineproof}

In the following, we present the average sum rate and the corresponding effective gain of VCC under MF precoding with imperfect CSIT. We recall that the power control parameters $\alpha^2$ and $\alpha_0^2$ have been derived in Proposition~\ref{alpha_prop}, while $\Xi_1$ and $\Xi_2$ are presented in Proposition~\ref{Xi_prop}.  %\nocite{7959863}
\begin{theorem}\label{VCC_thm}
   Under MF precoding and in the presence of imperfect CSIT, the average sum rate $ \bar R_{\rm sum}  \triangleq \mathbb{E}\{  R_{\rm sum} \}$ can be approximated as
    \begin{align}\label{sumRate_ave}
        \bar R_{\rm sum} \!\approx\!  \xi_{G,Q} GQ  \log_2\!\left( 1 +  \frac{ \alpha^2 \big(\Xi_1 + \sigma_e^2 L (2\beta+\Omega) \big) }{ 1 + \alpha^2  (Q-1) \Xi_2 } \right),
    \end{align}
   and the effective gain can be approximated as
    \begin{align}\label{gain_thm}
         &\mathcal{G} \notag\\
         &\approx \! \frac{ \max\limits_{Q \in \{2,\cdots,Q_{\max}\}}  GQ \xi_{G,Q} \log_2\left( \! 1 +  \frac{ \alpha^2 \big(\Xi_1 + \sigma_e^2 L (2\beta+\Omega) \big) }{ 1 + \alpha^2  (Q-1) \Xi_2 } \right)  }{  \max\limits_{Q' \in \{2,\cdots,Q_{\max}'\}}  Q'\xi_{1,Q'} \log_2\left( \! 1 +  \frac{ \alpha_0^2 \big(\Xi_1 + \sigma_e^2 L (2\beta+\Omega) \big) }{ 1 + \alpha_0^2  (Q'-1) \Xi_2 } \right) }.
    \end{align}
\end{theorem}

\begin{inlineproof}
The proof is relegated to Appendix~\ref{VCC_thm_proof}.
\end{inlineproof} 

\begin{remark}
   Our analytical method differs from conventional approaches (e.g., \cite{Miao_TAES,Kshitija}) that also adopt Rician-shadowed fading for modeling terrestrial-satellite channels. In those studies, the typical procedure involves first expressing the performance metrics (e.g., average capacity) as infinite integrals derived from the probability density function (PDF) of the channel gain, which are then evaluated or approximated to obtain closed-form results. In contrast, our approach avoids direct integration over the PDF and instead computes a few simple expectations. This not only improves tractability and yields a compact expression for parameterizing the delivery performance of VCC in SATCOM, but also ensures that the  approximation becomes increasingly tight as $L$ and $Q$ grow.
\end{remark}

\begin{remark}\label{MF_robust_rem}
In the average sum rate expression \(\bar R_{\rm sum}\) in~\eqref{sumRate_ave}, the intra-group interference term  is
\[
\alpha^{2}(Q-1)\,\Xi_{2}
=\frac{P_t\,(Q-1)}{G\,Q\,(2\beta+\Omega)},
\]
which is independent of the CSIT error, and which underscores the robustness of MF precoding.
\end{remark}

%%%%%%%%%%%%%%%%%%%%%%%%%%%%%%%%%%%%%%%%%%
\section{Numerical Performance Evaluation and Validation}\label{numerical_sec}
%%%%%%%%%%%%%%%%%%%%%%%%%%%%%%%%%%%%%%%%%%

\renewcommand*{\arraystretch}{1.35}%
\begin{table}[t!]
  \centering~
  \captionsetup{font={footnotesize}}
  \caption{Parameters in three typical shadowing scenarios \cite{Abdi}}\label{Scenario_Tab}
   \scalebox{0.95}{
  \begin{tabular}{ c| c c c }
  \hline
  Shadowing Scenarios  & $m$ & $\beta$ & $\Omega$ \\
  \hline  		
  Frequent Heavy Shadowing (FHS) & 0.739 & 0.063 & $8.97 \times 10^{-4}$ \\
  Average Shadowing (AS) & \;\;10.1\; & 0.126 & 0.835 \\
 Infrequent Light Shadowing (ILS) & 19.4 & 0.158 & 1.29 \\
 \hline
  \end{tabular}}~\vspace{-0.2cm}
\end{table}
\renewcommand*{\arraystretch}{1}%

   \begin{figure}[!t]
             \centering
             \includegraphics[width= 3.5 in]{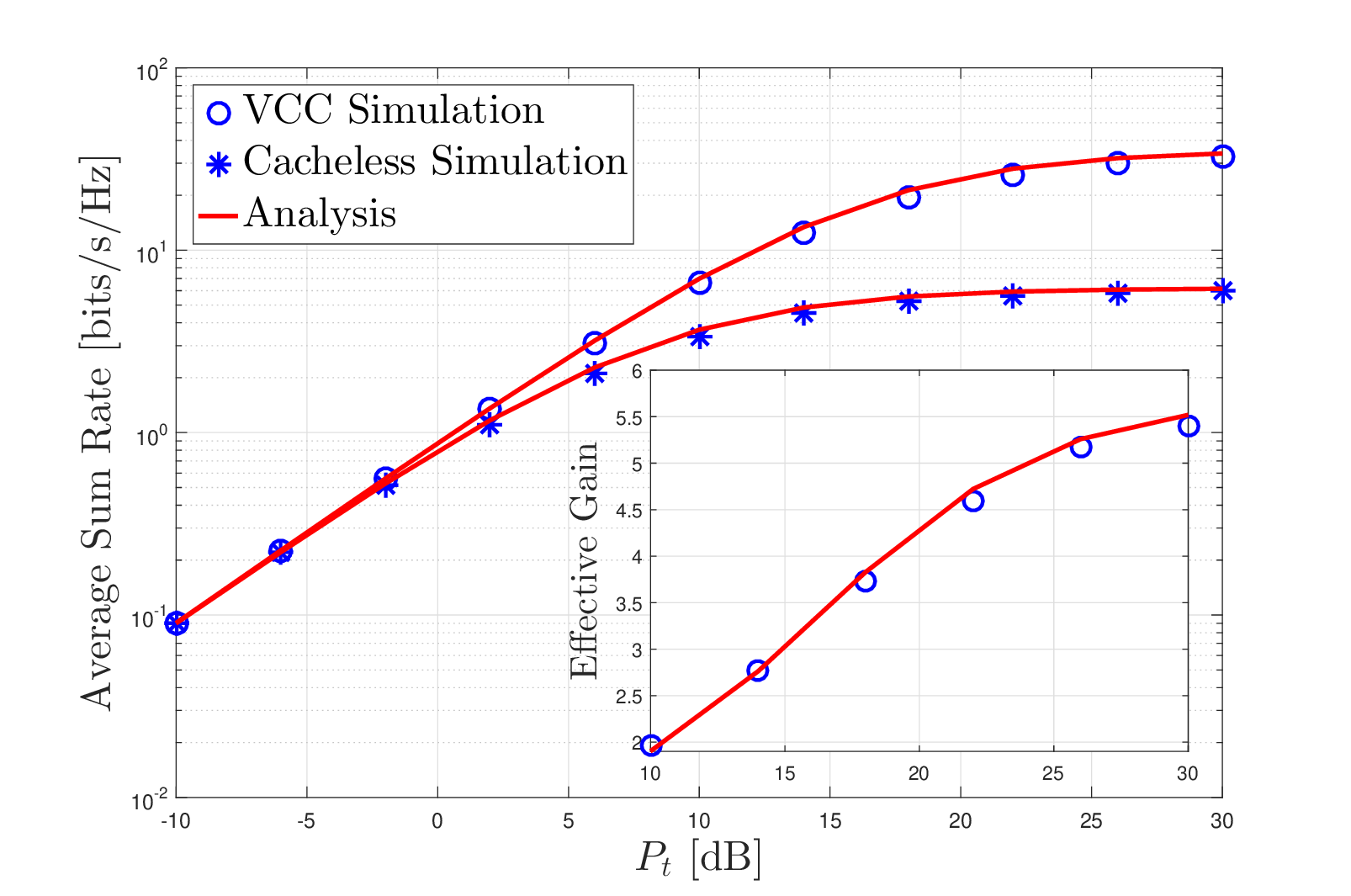}
                \vspace{-0.5cm}
             \captionsetup{font={footnotesize}}
		    \caption{Average sum rate and effective gain versus $P_t$ for $L=8$ and $Q_{\max} = Q_{\max}' = 8$ in FHS, where $\text{SNR}_{\text{ave}} = P_t - 9.0 $ [dB]. %This pertains mainly to GEO settings, and is thus less related to VoD applications. 
            %. where the average downlink SNR at each user is $P_t - 9.0 $ [dB].
            }\vspace{-0.3cm}\label{FHS_plot}
        \end{figure}

      \begin{figure}[!t]
             \centering
             \includegraphics[width= 3.5 in]{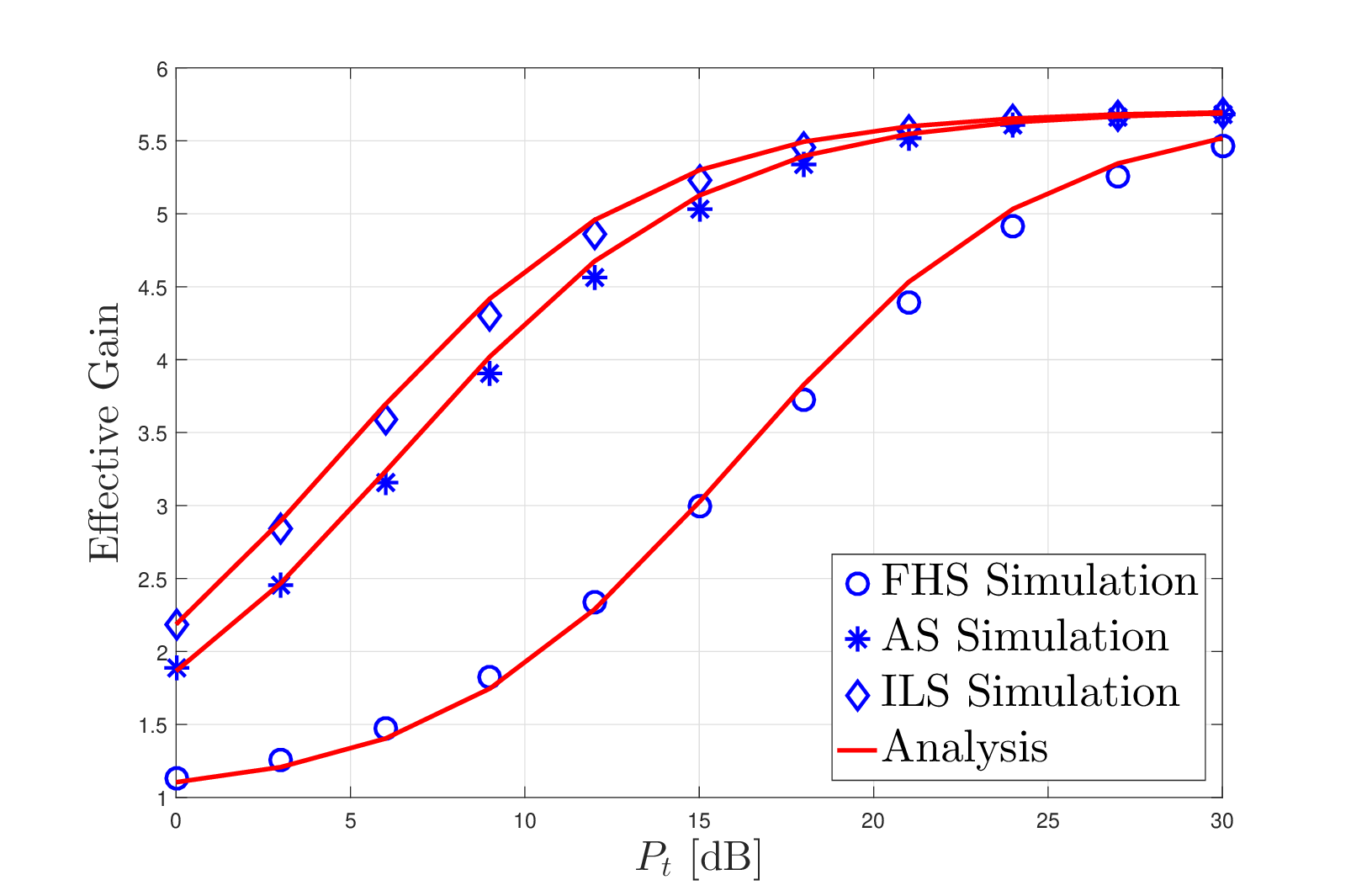}  \vspace{-0.5cm}
             \captionsetup{font={footnotesize}}
		    \caption{Effective gain versus $P_t$ for $L=8$ and $Q_{\max}=Q_{\max}'=8$, where the values of $\text{SNR}_{\text{ave}}$ in FHS, AS and ILS are respectively $P_t - 9.0 $ [dB], $P_t + 0.4$ [dB], and $P_t + 2.1$ [dB].}\vspace{-0.3cm}\label{VariousShadow_plot}
        \end{figure}

In this section, we numerically evaluate and plot the performance of VCC in SATCOM systems. %These simulations provide insights into the effective coded caching gain and the average sum rate under various shadowing scenarios and system configurations.
Unless stated otherwise, the variance of the estimation error $\sigma_e^2$ is assumed to be $\frac{1}{8}$ of the AWGN power. We consider low-mobility land terminals, and therefore assume a coherence length of  \( T = 10^4 \), which can correspond to a coherence time of 10~ms and a coherence bandwidth of 1~MHz.\footnote{As a practical reference, for GEO satellite–ground channels, the coherence time can reach approximately 500 ms after excluding system overheads such as gateway and user terminal processing delays \cite{8353925}. For a 600 km-altitude LEO orbit, the satellite–ground round-trip time (RTT) is about 4 ms. Given dense LEO satellite deployment and precise antenna alignment, the LEO-to-ground channel can exhibit a coherence time of 10 ms even when accounting for the  RTT and system overheads. Moreover, a 1 MHz coherence bandwidth is reasonable for LEO-to-ground links under conditions with a strong LOS component (e.g., medium and light shadowing).} Moreover,  we also assume \( G = 6 \) and \( \Theta = 12 \). The choice of {\color{black}{\( G = \Lambda \gamma +1 = 6 \)}} reflects the achievable theoretical gain under practical file size limitations, as discussed in \cite{Zhao_VectorCC}. {\color{black}{This value can arise from multiple combinations of the number of cache states $\Lambda$ and the cache size $\gamma$, such as $\Lambda=50$ with $\gamma=\frac{1}{10}$ or $\Lambda=80$ with $\gamma=\frac{1}{16}$.}}
%Consequently, the variance of the estimation error is \( \sigma_e^2 = \frac{1}{\Theta_1} = \frac{1}{4} \) assuming an uplink SNR of 0 dB for channel estimation at the satellite gateway \cite{Arti_NCC}. 

\begin{figure}[!t]
             \centering
             \includegraphics[width= 3.5 in]{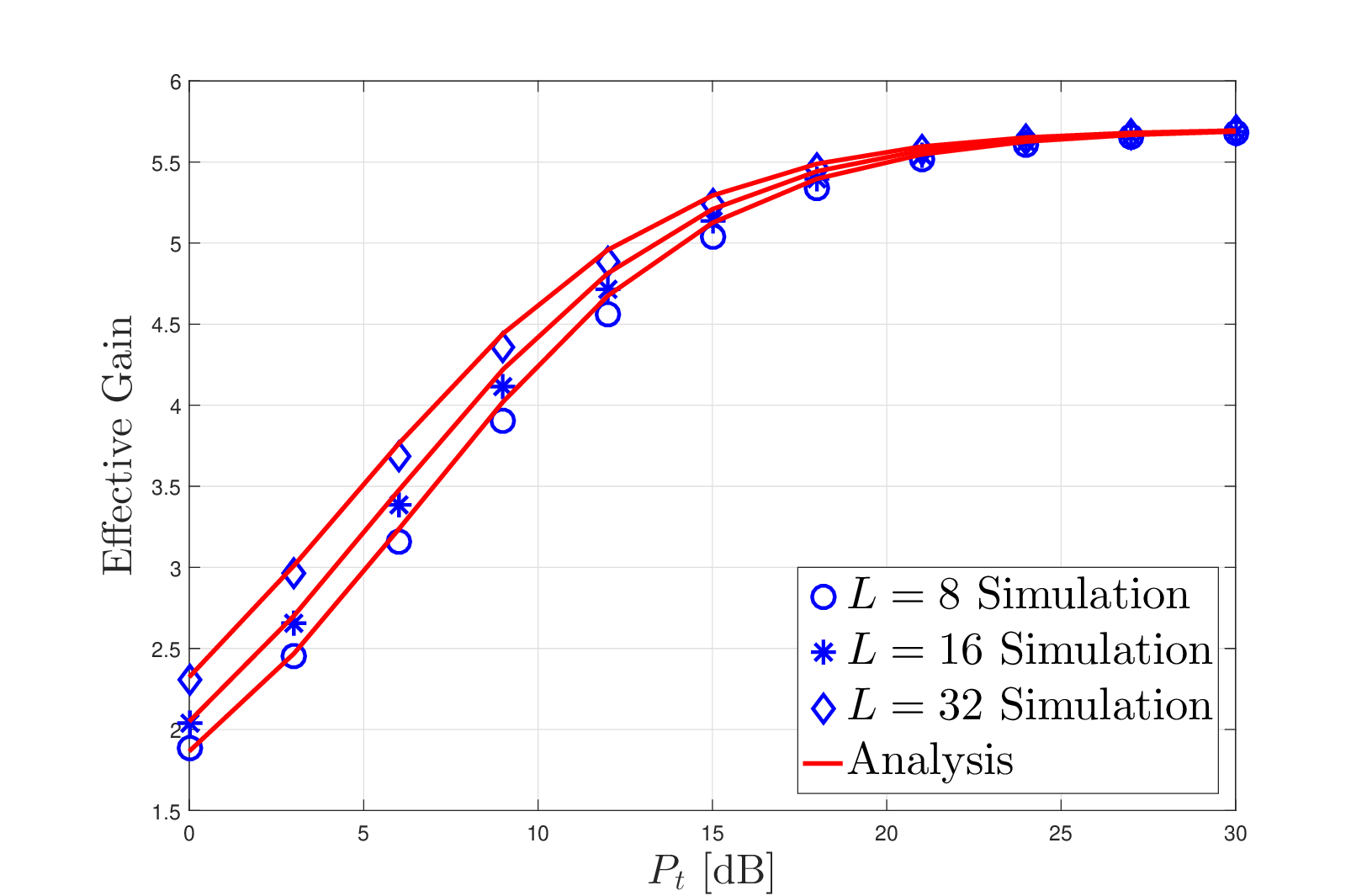}  \vspace{-0.5cm}
             \captionsetup{font={footnotesize}}
		    \caption{Effective gain versus $P_t$ in AS, where $Q_{\max}=Q_{\max}' = 8$ and $\text{SNR}_{\text{ave}} = P_t + 0.4$ [dB]. For $P_t \approx 18.1$ dB in AS, the $\text{SNR}_{\rm ave} \approx 18.5$ dB, corresponding to gain $\ge 5$.\\            
          %  Effective gain versus $P_t$ in AS, where the average downlink SNR at each user is $P_t + 0.4$ [dB].
        }\vspace{-0.5cm}\label{TransmitAN_plot}
        \end{figure}

\begin{figure}[!t]
             \centering
             \includegraphics[width= 3.5 in]{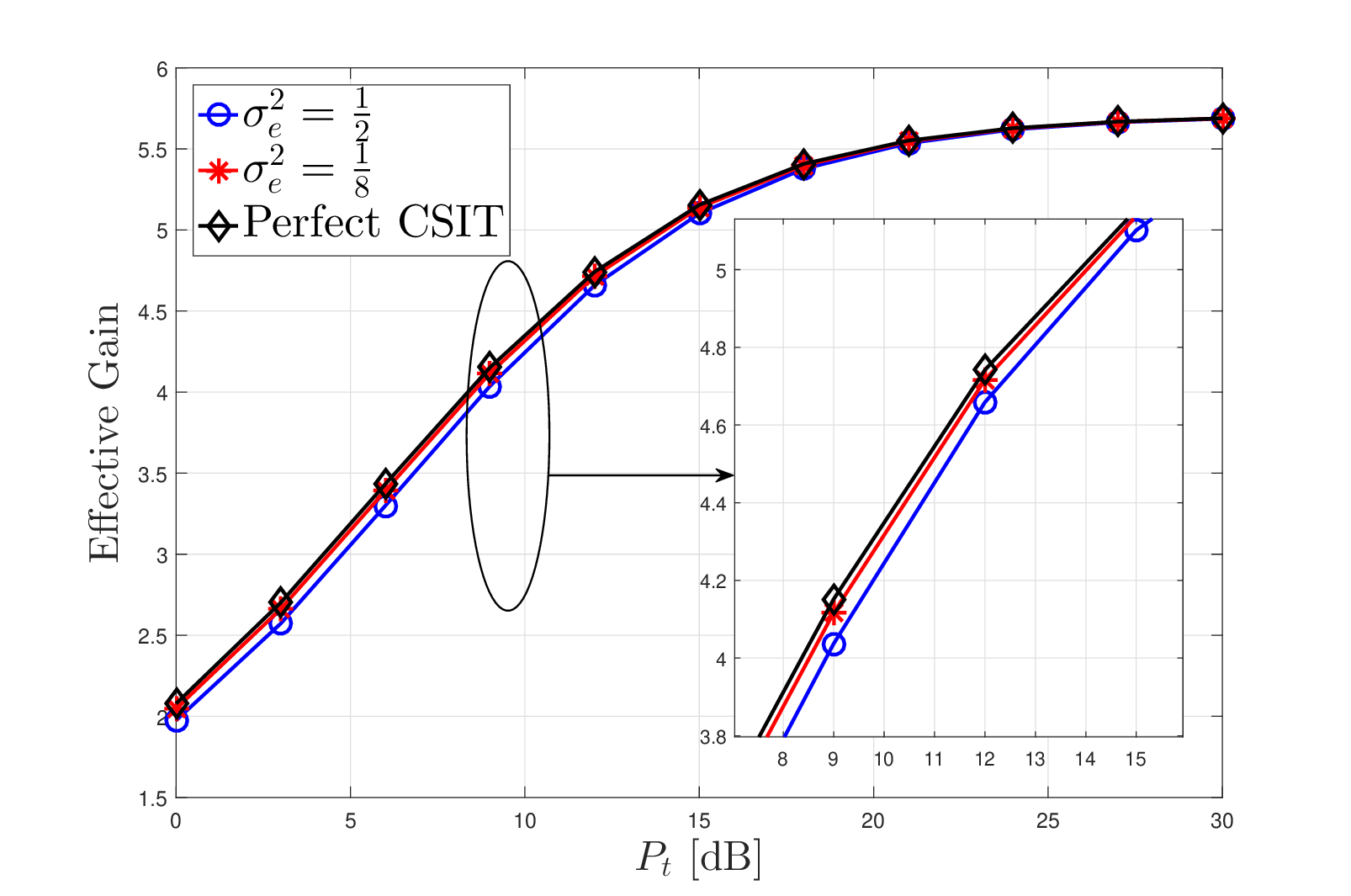}  \vspace{-0.5cm}
             \captionsetup{font={footnotesize}}
		    \caption{Effective gain versus $P_t$ in AS, where $L=16$,  $Q_{\max}=Q_{\max}' = 8$, and $\text{SNR}_{\text{ave}} = P_t + 0.4$ [dB]. For $P_t \approx 18.1$ dB in AS, the $\text{SNR}_{\rm ave} \approx 18.5$ dB, corresponding to gain $\ge 5$.
          %  Effective gain versus $P_t$ in AS, where the average downlink SNR at each user is $P_t + 0.4$ [dB].
        }\label{CSIT_plot}\vspace{-0.3cm}
        \end{figure}

\begin{figure}[!t]
             \centering
             \includegraphics[width= 3.5 in]{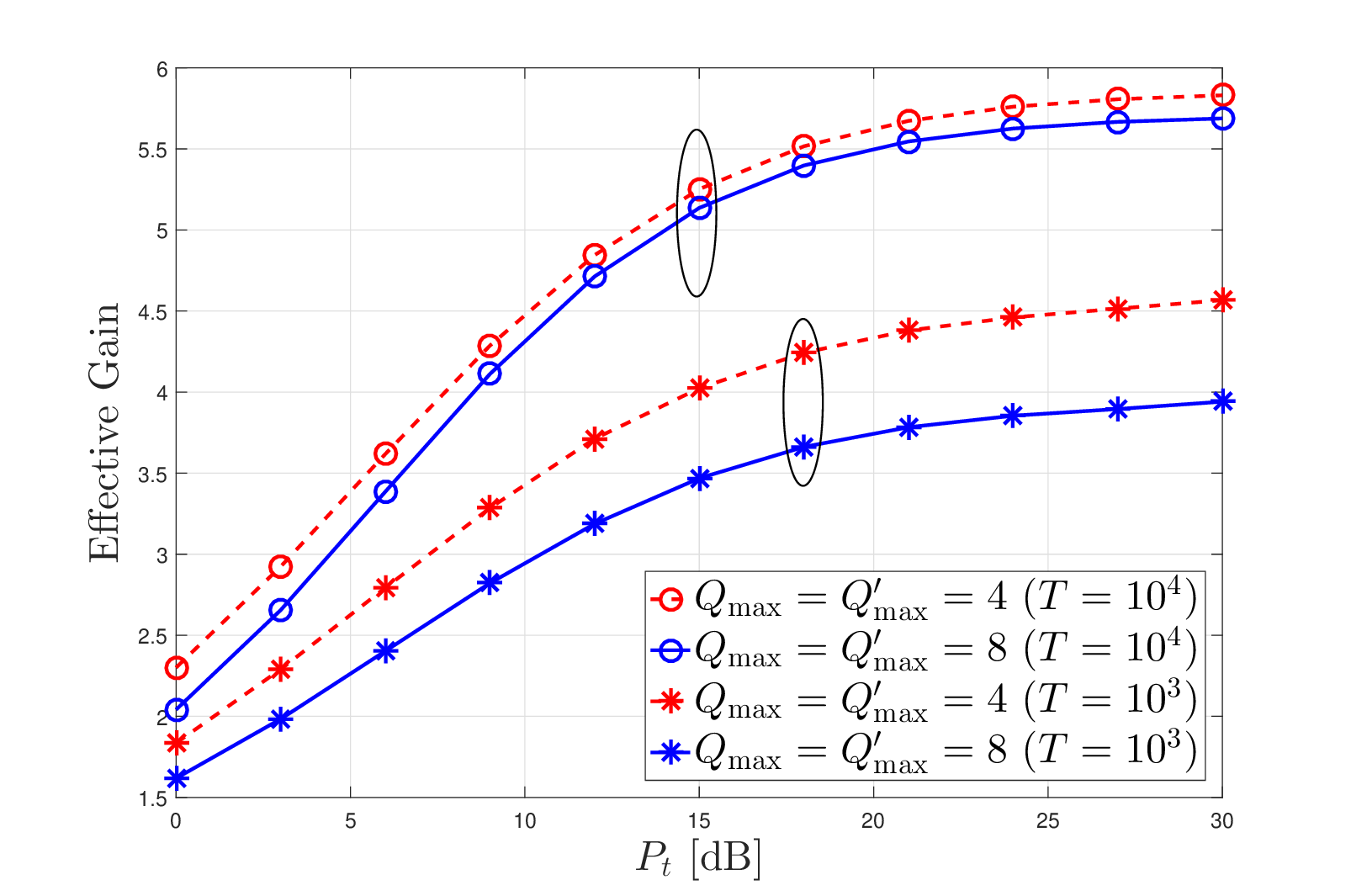}  \vspace{-0.5cm}
             \captionsetup{font={footnotesize}}
		    \caption{Effective gain versus $P_t$ in AS, where $L=16$ and $\text{SNR}_{\text{ave}} = P_t + 0.4$ [dB]. For $P_t \approx 18.1$ dB in AS, the $\text{SNR}_{\rm ave} \approx 18.5$ dB, corresponding to gains between $3.5$ and $5.5$.\\
          %  Effective gain versus $P_t$ in AS, where the average downlink SNR at each user is $P_t + 0.4$ [dB].
        }\vspace{-0.3cm}\label{T3Vs4_fig}
        \end{figure}

{\color{black}{\subsection{Numerical Results Under Static Channel}

In this subsection, we consider the static channel model as analyzed in this paper.}}
We evaluate the system performance under three different shadowing scenarios, with the specific Rician-shadowed channel parameters listed in Table~\ref{Scenario_Tab}. 
%We note that light-to-moderate shadowing occurs more frequently in LEO-to-ground transmissions, whereas GEO-to-ground links are more often subject to  heavy shadowing \cite{Li_VT}.
According to \cite[Eq.~(5.3)]{EURECOM+7083}, the average downlink SNR over a Rician-shadowed channel is \(\text{SNR}_{\text{ave}} = P_t (2\beta + \Omega)\), measured from a single-antenna satellite to a user terminal. %It is worth mentioning that $\text{SNR}_{\text{ave}}$ can reach about 15--20~dB for a 600~km-altitude LEO satellite under light shadowing conditions \cite{Eunsun,3GPP_TR_38_811_v15_3_0}. 
For reference,  $\text{SNR}_{\text{ave}}$ is also indicated in Figs.~\ref{FHS_plot}--\ref{T3Vs4_fig}.

We proceed to illustrate in Figs. \ref{FHS_plot}--\ref{T3Vs4_fig}, the spectral efficiency gains brought about by introducing VCC in a downlink VoD-delivery satellite system. This gain (y-axis) is a \emph{multiplicative} gain on spectral efficiency, and it incorporates the differing CSIT overhead and CSIT estimation errors of the two compared systems. We also recall that the recorded gains are over {\color{black}{downlink}} systems that enjoy the same resources (same power, same number of antennas), and which are \emph{independently} optimized with respect to the multiplexing–vs-beamforming tradeoff. Finally, lets also recall from Remark~\ref{rem:NoCheat} that the gains do not arise from pre-caching, but that indeed these are gains on the speed of VoD delivery of data that must actually be delivered after the demands are declared.

Fig.~\ref{FHS_plot} illustrates the average sum rate and the effective gain versus \( P_t \) in the most challenging shadowing environment, frequent heavy shadowing (FHS).
%, which as suggested relates mainly to GEO settings, and thus not directly related to VoD applications. 
{\color{black}{We note that FHS is not limited to GEO satellite–terrestrial links and can also arise in LEO scenarios \cite{Eunsun,Fangxin,Shizhao}.}}
As expected, the average sum rate increases with \( P_t \), and when we reach $P_t =  15$ dB, VCC yields a gain---over a system with the same power and antenna resources, but without VCC---of $\times 3$ in spectral efficiency (this is also often referred to as a 200\% boost in spectral efficiency).  What we also note is that the analytical outcomes from Theorem~\ref{VCC_thm} (solid lines) are tightly validated by simulations.

%. Even under FHS conditions, VCC still achieves a notable performance improvement.  For instance, when

Fig.~\ref{VariousShadow_plot} shows the effective gain versus \( P_t \) across various shadowing conditions: FHS, average shadowing (AS), and infrequently light shadowing (ILS). For the three shadowing cases examined, VCC delivers sizable spectral efficiency gains relative to the cacheless system in the practical $P_t$ regime. For instance, according to the link budget in Table~\ref{budget_tab} 
($P_t \approx 18.1$ dB, corresponding to an EIRP\footnote{EIRP (Effective Isotropic Radiated Power) is defined as transmit power (in dBW) plus antenna gain (dBi) minus feeder losses (dB). It represents the apparent radiated power in the direction of maximum gain, as if from an ideal isotropic antenna. For instance, a satellite with 10 W transmit power (10 dBW) and 30 dBi antenna gain yields an EIRP of 40 dBW.} of $45$ dBW), the spectral efficiency is boosted by more 
than $400\%$ (i.e., $\times 5$) in both AS and ILS cases. Even under 
the most adverse FHS condition, VCC can still \emph{quadruple} the throughput compared to its cacheless counterpart.

\begin{example}
To clarify the above points, let us revisit the example of Fig.~\ref{VariousShadow_plot}, focusing on the curve corresponding to the AS case. Consider a satellite equipped with $L=8$ transmit antennas, and let $P_t = 9 \text{ dB}$. For this setting, the plot shows a multiplicative gain of $\times 4$.

What does this mean exactly? Suppose an operator, with the above resources ($L=8$ transmit antennas and $P_t = 9\text{ dB}$), optimizes its system under the MF precoding assumption by balancing multiplexing and beamforming gains. After accounting for CSIT overhead, the resulting spectral efficiency is $R_1$ bits/s/Hz. Now, applying the VCC approach with the same resources $L,P_t$, and after accounting for the new CSIT overhead, yields a spectral efficiency $R_{\text{vcc}}$. Our results show that 
$R_{\text{vcc}} \approx   4 \times R_1.$  To further clarify, recall Remark~\ref{rem:NoCheat}, which effectively states that the recorded gain persists even if the baseline downlink system (without VCC) is equipped with the same receiver-side cache as VCC. In that case, the baseline would indeed avoid transmitting already cached content, yet the multiplicative advantage of VCC over the baseline remains unchanged.
\end{example}

%{\color{red}{
%In practice, the beamwidth resolution of a satellite spot beam limits the number of users that can be simultaneously served with the same time-frequency resources. Hence, in Fig.~\ref{GainQ8_fig} this constraint is incorporated by restricting the number of precoded users to eight, i.e., $Q,Q' \leq 8$. Compared to Fig.~\ref{VariousShadow_plot}, the effective gain in the AS and ILS shadowing scenarios remains almost unchanged, whereas in the adverse FHS case it increases significantly for $P_t \ge 5$ dB. }}

Fig.~\ref{TransmitAN_plot} plots how the effective gain varies with the number of transmit antennas $L$. This aspect is particularly relevant in satellite systems, where large-scale antenna arrays have strong potential for deployment on LEO satellites~\cite{Bhavani,9110855}. The main observation here is that having more transmit antennas can in fact amplify the gain from VCC over the baseline system (with the same increased $L$). This observation---that higher $L$ yields higher multiplicative gains---marks a difference from the case of terrestrial networks over Rayleigh fading channels (cf. \cite[Fig.~3]{Zhao_VectorCC}), and this difference can be attributed to the fundamentally different nature of the channels and the spatial resolution constraint in SATCOM.

%Fig.~\ref{TransmitAN_plot} also illustrates the effect of CSIT imperfections, revealing a negligible performance penalty due to CSIT errors with modest power. This robustness is partially attributed to the use of MF precoding, but also to the fact that VCC cross-group interference cancellation does not need CSIT (for more on this, see also~\cite{Zhao_VCC_TWC2}).
%Overall, these results highlight the substantial advantages of vector coded caching in SATCOM systems, even under realistic conditions with varying shadowing and imperfect CSIT.

Fig.~\ref{CSIT_plot} depicts the sensitivity to CSIT imperfection. Increasing the CSIT-estimation error slightly reduces the effective gain; the reduction is marginal across the entire \(P_t\) range, highlighting that, with MF precoding, the inter-user interference term in the SINR is nearly insensitive to CSIT errors (see Remark~\ref{MF_robust_rem}). In contrast, the ZF-based results in \cite{Zhao_VCC_TWC2} show an \emph{increase} in effective gain as the CSIT error grows, because VCC’s cross-group interference cancellation is CSIT-free and therefore comparatively insensitive to CSIT errors, causing the relative-gain metric to rise with the error level.

    \renewcommand*{\arraystretch}{1.35}%
\begin{table}[!t]
  \centering~
  \captionsetup{font={footnotesize}}
  \caption{Link budget evaluation for a single-antenna LEO satellite at 600 km altitude \cite{Eunsun,3GPP_TR_38_811_v15_3_0}, with EIRP = 45 dBW, user terminal G/T = 5 dB/K (small VSAT), carrier frequency 20 GHz, bandwidth 36 MHz, zenith angle (90°), including gaseous absorption (0.9 dB) and rain attenuation (10 dB). The resulting $P_t$ is approximately 18.1 dB.}\label{budget_tab}
   \scalebox{0.95}{
  \begin{tabular}{ c| c c c }
  \hline
  Shadowing Scenarios  & FHS & AS & ILS \\
  \hline
  $\text{SNR}_{\rm ave}$ [dB] &  $9.1$  & $18.5$ & $20.2$ \\
  Effective Gain in Fig.~\ref{VariousShadow_plot}  & $\ge 4$ &  $\ge 5$  & $\ge 5.5$\\
  \hline
  \end{tabular}}~\vspace{-0.3cm}
\end{table}
\renewcommand*{\arraystretch}{1}%

The size of the coherence block $T$ has a direct impact on the performance of VCC. With a fixed CSI acquisition cost, a larger $T$ reduces the relative overhead and thereby increases the effective throughput, which translates into a higher effective gain. Conversely, smaller $T$ values amplify the impact of CSI cost and degrade the gain. Fig.~\ref{T3Vs4_fig} illustrates the effective gain achieved for different values of $T$.  As \(T\) decreases, the gain drops markedly, while the dependence on \(T\) is less pronounced at low $P_t$. When \(T\) is large (e.g., \(10^4\)), the choice of \(Q_{\max},Q'_{\max}\) has little influence; for small \(T\) (e.g., \(10^3\)), increasing the \emph{spatial-multiplexing cap}—i.e., the maximum number of simultaneous spatial streams afforded by multiple antennas—actually reduces the effective gain, while a smaller cap increases it. It is also worth noting that even with \(T=10^3\), the gains remain substantial: at \(P_t=18.1\)~dB, both \(Q_{\max},Q'_{\max}\in\{4,8\}\) yield an effective gain above $3.5$, i.e., more than a \emph{three-and-a-half–fold} throughput improvement over conventional MU-MIMO SATCOM.
%It is evident that reducing $T$ significantly decreases the effective gain, although this effect is less pronounced at low SNR. %However, in the low-SNR regime, the effective gain is typically below one, which already indicates that conventional (cacheless) MU-MIMO transmission would be more favorable.
% It is also worth noting that even with $T=500$, the effective gain still reaches approximately $4$ at $P_t=15$~dB, corresponding to \emph{quadrupling} the throughput compared to conventional MU-MIMO in SATCOM.

%These numerical results, in addition to revealing that VCC achieves ~3×–5× more bits/Hz than conventional in Satellite MU-MISO, also suggest a broader design insight: when aiming to meet a given spectral efficiency requirement, VCC can achieve the same performance with significantly fewer PHY resources --- such as transmit antennas, RF chains, or accurate CSIT --- compared to traditional MU-MISO (multi-beam) SATCOM systems. This opens up new opportunities for designing more cost-effective and resource-efficient SATCOM architectures.
VCC’s $3\times$--$5\times$ spectral efficiency gain not only outperforms conventional MU-MISO, but also shows that similar performance can be reached with fewer antennas, RF chains, or less accurate CSIT---opening new research directions in designing leaner, more efficient SATCOM architectures.

\begin{figure}[!t]
             \centering
             \includegraphics[width= 3.5 in]{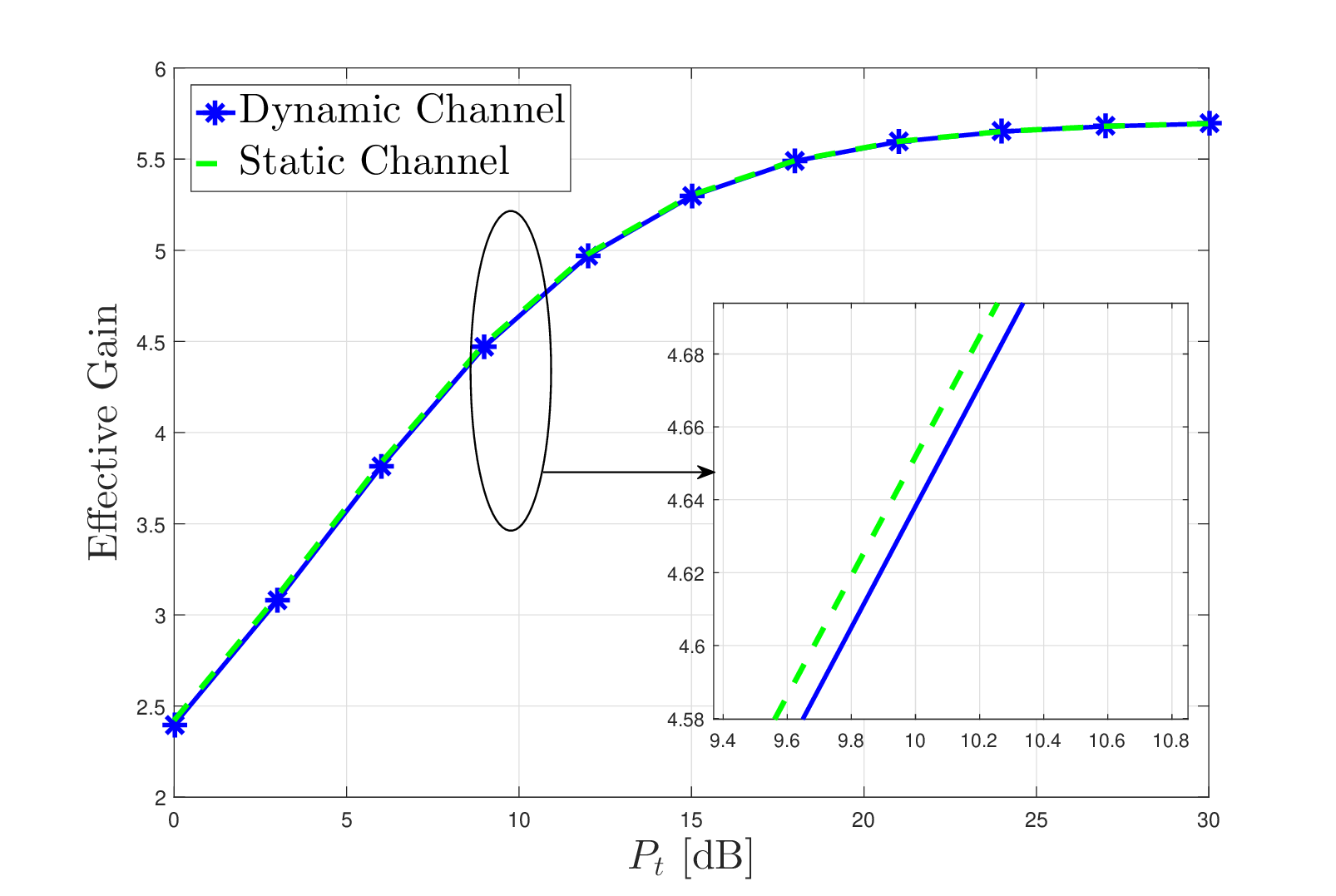}  \vspace{-0.5cm}
             \captionsetup{font={footnotesize}}
		    \caption{Effective gain versus $P_t$ in various channel models, where $L=16$, $Q_{\max}=Q_{\max}'=8$, $D=10$ km and $\eta = 0.35$ (urban area).\\
          %  Effective gain versus $P_t$ in AS, where the average downlink SNR at each user is $P_t + 0.4$ [dB].
        }\vspace{-0.5cm}\label{Gain_Dyna_fig}
        \end{figure}

{\color{black}{
\subsection{Extension to Dynamic Channel}
In this subsection, we consider a dynamic channel model for SATCOM in order to demonstrate that the multiplicative spectral-efficiency gains achieved by VCC over the cacheless MU-MISO baseline are preserved under more practical channel conditions.
We consider users that are \emph{randomly and uniformly} distributed over a terrestrial coverage area with a radius $D$. The expectation in the effective gain definition in~\eqref{Gain_def} is further averaged over the random realizations of the user spatial distribution. We do not explicitly model Doppler effects arising from the relative motion between the serving satellite and terrestrial terminals. In practical deployments, such effects can be accurately estimated and effectively compensated, since satellite orbital parameters and relative user locations are typically known or can be reliably tracked \cite{Eunsun,Li_VT}.

Owing to the spatial distribution of users, the satellite--terrestrial channels experience rapid state variations between LOS and NLOS conditions. Following~\cite{Fangxin,Shizhao}, the dynamic transitions between LOS and NLOS conditions are modeled using a \emph{two-state Markov process}.
Specifically, under the dynamic channel model, the channel vector $\mathbf{h}_k \in \mathbb{C}^{L \times 1}$ of the $k$-th user is given by
\begin{align}\label{h_dynamic}
\mathbf{h}_k =
\begin{cases}
\mathbf{h}_k^{\mathrm{LOS}}, & \text{with probability } \mathcal{P}_k, \\
\mathbf{h}_k^{\mathrm{NLOS}}, & \text{with probability } 1-\mathcal{P}_k,
\end{cases}
\end{align}
where $\mathcal{P}_k$ denotes the probability of LOS propagation determined by the elevation angle $\zeta_k$ between the $k$-th user and the serving satellite. As suggested in~\cite{Hourani,Fangxin}, this probability can be simply expressed as 
\begin{align}\label{P_k_LOS}
    \mathcal{P}_k = \exp\!\left(-\eta \cot \zeta_k\right),
\end{align}
where $\eta$ is an environment-dependent parameter.
The LOS and NLOS channel components, $\mathbf{h}_k^{\mathrm{LOS}}$ and $\mathbf{h}_k^{\mathrm{NLOS}}$ in \eqref{h_dynamic}, are assumed to be statistically independent and both follow the Rician-shadowed fading model proposed in~\cite{Abdi}. Specifically, we adopt the ILS model to characterize $\mathbf{h}_k^{\mathrm{LOS}}$ and the FHS model to characterize $\mathbf{h}_k^{\mathrm{NLOS}}$. The corresponding channel parameters are summarized in Table~\ref{Scenario_Tab}. We note that the analytical expression in~\eqref{P_k_LOS} for modeling the LOS probability has been shown to be in good agreement with the 3GPP model \cite{Hourani}.

Fig.~\ref{Gain_Dyna_fig} illustrates the effective gain achieved by VCC under static and dynamic channel models for a terrestrial coverage area with radius $D=10$~km, corresponding to a medium-sized city. The satellite altitude $H$ is set to 600~km, and the serving LEO satellite is located at the zenith position above the center of the terrestrial coverage area. Under this geometry,\footnote{{\color{black}{The pathloss variation across users is negligible. For instance, the maximum pathloss occurs at the cell-edge users located at a distance of $D=10$~km from the coverage center, whose distance to the serving satellite is $\sqrt{H^2 + D^2} \approx 600.08$~km. Consequently, for simplicity, we assume identical pathloss for all users.
}}} the elevation angle $\zeta_k$ of terrestrial terminals ranges from $89^\circ$ to $90^\circ$.
In the static case, the channel is modeled using the ILS scenario in Table~\ref{Scenario_Tab}, whereas the dynamic channel follows the model in~\eqref{h_dynamic}, where the parameter $\eta$ in \eqref{P_k_LOS} for $\mathcal{P}_k$ is set to $0.35$, corresponding to an urban environment~\cite{Hourani,Fangxin}. It is observed that the effective gain under dynamic channels closely matches that of the static case across the entire $P_t$ range, with only negligible performance degradation. This confirms that the multiplicative spectral-efficiency gains provided by VCC over the cacheless MU-MISO baseline are well preserved under dynamic satellite--terrestrial channels. It also provides justification for adopting a static and symmetric channel in the channel modeling and performance analysis of this work.
}}

%%%%%%%%%%%%%%%%%%%%%%%%%%%%%%%%%%%%%%%%%%%%%%%%%%%%%%%
\section{Conclusion and Discussions}\label{Conclusion_sec}
%%%%%%%%%%%%%%%%%%%%%%%%%%%%%%%%%%%%%%%%%%%%%%%%%%%%%%%

In this paper, we analyzed the performance of VCC in state-of-art (multi-beam) SATCOM systems, focusing on scenarios characterized by low-to-moderate SNR conditions and various shadowing environments. We derived an analytical model that accurately derives the average sum rate and the effective gain, validated by simulations under different system configurations and channel conditions.
Our results demonstrate that VCC provides substantial spectral efficiency gains even in challenging environments. %; even in the most challenging FHS environment, where the average received SNR is a mere 6 dB, VCC offers a 150\% boost in spectral efficiency, corresponding to a multiplicative boost of $\times 2.5$. 
In the LEO setting  experiencing average and light shadowing, the gains are more pronounced and VCC can boost the spectral efficiency by a stunning $300$--$550\%$, {\color{black}{even in more challenging dynamic channels.}}  %Moreover, we found that increasing the number of transmit antennas enhances the effective gain, and that the difference in performance between perfect and imperfect CSIT is negligible, highlighting the robustness of VCC in practical SATCOM applications.
%Overall, this study confirms that VCC is a promising technique for enhancing data delivery efficiency in SATCOM systems, particularly for low-mobility users in challenging shadowing conditions. 
%\nocite{9163148}
This work further positions VCC not just as a tool for extending SATCOM coverage to remote areas, but as a promising enabler for narrowing the significant spectral efficiency gap between satellite and fiber-optic communication systems for VoD.

%Future research can explore several avenues to further enhance the performance of VCC in SATCOM systems. First, developing advanced transmission schemes tailored to mitigate the severe effects of FHS could lead to improved spectral efficiency in challenging environments. Additionally, fully leveraging user-side caching to design specialized channel estimation techniques can significantly reduce the required pilot length. This reduction would further boost the effective gain by decreasing the CSI acquisition cost. These improvements could make VCC even more practical and efficient for future satellite communications.

A key factor affecting the performance gap between VCC and the traditional baseline is the CSI acquisition cost, which grows with the need to serve approximately \(G\) times more users. Our system model adopts a deliberately conservative training scheme: it omits advanced techniques like pilot reuse or coordination and assumes a fixed pilot length regardless of SNR. While this ensures robust, and perhaps conservative, performance estimates, it also indicates room for further gains through optimized training strategies---a direction we leave for future work. {\color{black}{Moreover, incorporating time-varying channel effects induced by satellite–user relative motion, as well as exploiting distance-based spatial separation to further suppress inter-user interference in coded caching–aided SATCOM delivery, constitute interesting directions for future research.}}

We advocate that VCC is a novel method for accelerating satellite video delivery, doing so by rethinking the precoding structure at the PHY---not by guessing user behavior or caching entire videos in advance. Our approach is a unicast approach that retains standard VoD semantics: content is requested interactively, and our system delivers it more efficiently using a redesigned transmission method. Specifically, we modify how transmit symbols are constructed at the baseband precoding level---with no changes to the RF front-end, modulation scheme, or antenna hardware. The result is a $3\times$--$5\times$ spectral efficiency gain, enabled by layering multiple unicast streams over the same spectrum and decoding them using structured receiver logic. The modest storage at each user terminal is not used to bypass transmission but instead supports the decoding process. %Critically, this design is deployable via software or firmware updates on both the satellite and user terminal sides --- especially in vertically integrated systems like Starlink, which support OTA updates and DSP-based baseband chains. Because the core waveform remains compatible and all physical constraints are respected, this solution offers high performance with low architectural disruption --- a compelling fit for modern, software-defined satellite networks. 

%%%%%%%%%%%%%%%%%%%%%%%%%%%%%%%%%%%%%%%%%%%%%%%%%%%%%%%%%%%%%%%
%%%%%%%%%%%%%%%%%%%%%%%%%%%%%%%%%%%%%%%%%%%%%%%%%%%%%%%%%%%%%%%
\appendices
\renewcommand{\thesectiondis}[2]{\Alph{section}:}
%%%%%%%%%%%%%%%%%%%%%%%%%%%%%%%%%%%%%%%%%%%%%%%%%%%%%%%%%%%%%%%
%%%%%%%%%%%%%%%%%%%%%%%%%%%%%%%%%%%%%%%%%%%%%%%%%%%%%%%%%%%%%%%
%%%%%%%%%%%%%%%%%%%%%%%%%%%%%%%%%%%%%%%%%%%%%%%%%%%%%%%%%%%%%%%%%%%%%%%%%%%%%%%%%%%%

\section{Proof of Proposition~\ref{alpha_prop}}\label{alpha_prop_proof}
Considering the transmit signal under MF precoding in \eqref{transmit_signal_MF} and the average transmit power constraint $P_t$, 
%we have that
%\begin{align}\label{alpha_exp}
%        \mathbb{E}&\{ ||{\bf x} ||^2 \} = \alpha^2 \mathbb{E} \bigg\{ \bigg( \sum_{\phi \in \Psi} {\bf s}_\phi^H \hat{\bf H}_\phi \bigg)  \bigg( \sum_{\psi \in \Psi}  \hat {\bf H}_\psi^H {\bf s}_\psi \bigg)  \bigg\} = P_t \notag\\
    %    &\Longrightarrow \alpha^2 \sum_{\phi \in \Psi} \sum_{\psi \in \Psi}  \mathbb{E} \Big\{ {\rm Tr} \Big\{  {\bf s}_\phi^H \hat{\bf H}_\phi \hat{\bf H}_\psi^H {\bf s}_\psi \Big\}  \Big\} = P_t \notag\\
%        & \Longrightarrow \alpha^2 \sum_{\phi \in \Psi} \sum_{\psi \in \Psi}  \mathbb{E} \Big\{ {\rm Tr} \Big\{   \hat{\bf H}_\phi \hat{\bf H}_\psi^H {\bf s}_\psi  {\bf s}_\phi^H \Big\} \Big\} = P_t \notag\\
%         & \overset{(a)}{\Longrightarrow} \alpha^2 \sum_{\phi \in \Psi} \sum_{\psi \in \Psi} {\rm Tr}    \Big\{ \mathbb{E} \Big\{   \hat{\bf H}_\phi \hat{\bf H}_\psi^H \Big\} \mathbb{E} \Big\{  {\bf s}_\psi  {\bf s}_\phi^H \Big\} \Big\} = P_t  
     %^    &  \Longrightarrow \alpha^2 = \frac{P_t}{ \sum_{\phi \in \Psi}  {\rm Tr}    \bigg\{ \mathbb{E} \bigg\{   {\bf H}_\phi {\bf H}_\phi^H \bigg\} \bigg\} + \alpha^2 \sum_{\phi \in \Psi} \sum_{\psi \in \Psi \setminus \{\phi\} } {\rm Tr}    \bigg\{ \mathbb{E} \bigg\{   {\bf H}_\phi {\bf H}_\psi^H \bigg\} \bigg\} }
%\end{align}
%where $(a)$ follows from the interchange between the expectation and the trace operator. 
we can write the squared power control factor $\alpha^2$ as
\begin{align}\label{alpha2_exp}
    \alpha^2 = \frac{P_t}{ \sum_{\phi \in \Psi} \sum_{\psi \in \Psi} {\rm Tr}    \big\{ \mathbb{E} \big\{  \hat {\bf H}_\phi \hat {\bf H}_\psi^H \big\}  \big\}  }.
\end{align}
For $\phi= \psi$, we have that
\begin{align}\label{term1_alpha}
   \sum_{\phi \in \Psi}  {\rm Tr}    \big\{ \mathbb{E} \big\{  \hat {\bf H}_\phi \hat {\bf H}_\phi^H \big\} \big\}  &=  \sum_{\phi \in \Psi} \sum_{b \in [Q]} \mathbb{E} \{  ||\hat{\bf h}_{\phi,b}||^2 \}  \notag\\
   &= \sum_{\phi \in \Psi} \sum_{b \in [Q]} L (2 \beta + \sigma_e^2 + \Omega) \notag\\
   & = GQ L (2 \beta + \sigma_e^2 + \Omega) ,
\end{align}
where we consider the fact that $\mathbb{E}\{ ||\hat{\bf h}_{\phi,b}||^2 \} = L (2 \beta + \sigma_e^2 + \Omega)$ (cf. \cite[Prop. 5.1]{EURECOM+7083}).
Next, we consider the case of $\psi \neq \phi$, which yields that 
\begin{align}\label{term2_alpha}
    &\sum\nolimits_{\phi \in \Psi} \sum\nolimits_{\psi \in \Psi \setminus \{\phi\} } \!\!\! {\rm Tr}    \big\{ \mathbb{E} \big\{  \hat {\bf H}_\phi \hat {\bf H}_\psi^H \big\} \big\} \notag\\
    & = \sum\nolimits_{\phi \in \Psi} \sum\nolimits_{\psi \in \Psi \setminus \{\phi\} } \!\!\! {\rm Tr}    \big\{ \mathbb{E} \big\{  \hat {\bf H}_\phi \big\} \mathbb{E}\big\{ \hat {\bf H}_\psi^H \big\} \big\}
    \overset{(a)}{=} 0,
\end{align}
where \( (a) \) follows from the fact that, for a scalar channel gain ${h}_{\psi,b}^{(\ell)}$ in Rician-shadowed fading channels, the expectations of both the scatter components and estimation errors (each being complex Gaussian with zero-mean) are zero. Thus, only the LOS component with Nakagami–\(m\) amplitude needs to be modeled, whose mean is
\begin{align}
    \mathbb{E} \left\{ {h}_{\psi,b}^{(\ell)} \right\} = \frac{\Gamma\!\left(m+\tfrac{1}{2}\right)}{\Gamma(m)}
\sqrt{\frac{\Omega}{m}} \mathbb{E} \left\{ \exp\Big(\jmath \theta_{\psi,b}^{(\ell)} \Big) \right\},
\end{align}
where $\theta_{\psi,b}^{(\ell)}$ is the phase of the LOS component. 
As $\theta_{\psi,b}^{(\ell)}$ is uniformly distributed across channel blocks, we have
$\mathbb{E}\big\{\exp\big( \jmath \theta_{\psi,b}^{(\ell)}  \big) \big\} = 0$.
Consequently, the LOS phasor averages out, which leads to
\begin{align}
    \mathbb{E} \left\{ {h}_{\psi,b}^{(\ell)} \right\} =0.
\end{align}

Finally, substituting \eqref{term1_alpha} and \eqref{term2_alpha} into \eqref{alpha2_exp}, we can derive the closed-form expression for $\alpha^2$ in Proposition~\ref{alpha_prop}.

\section{Proof of Proposition~\ref{Xi_prop}}\label{Xi_prop_proof}
We first derive the expression for $\Xi_1$ as given in \eqref{X11_prop_eq}. Based on its definition in \eqref{Xi_def_eq}, $\Xi_1$ can be reformulated as
\begin{align}
    \Xi_1 &= \mathbb{E}\{ ||{\bf h}_{\psi,b}||^4 \}
    = \mathbb{E}\{ ||{\bf h}_{\psi,b}||^2 ||{\bf h}_{\psi,b}||^2 \} \notag\\
    &= \mathbb{E} \left\{ \bigg( \sum_{\ell=1}^L \big| {h}_{\psi,b}^{(\ell)} \big|^2  \bigg) 
 \bigg( \sum_{\vartheta=1}^L \big|{h}_{\psi,b}^{(\vartheta)}\big|^2 \bigg) \right\} ,
% &= \mathbb{E} \left\{ \sum_{\ell=1}^L  \sum_{\vartheta=1}^L \big| {h}_{\psi,b}^{(\ell)} \big|^2  \big|{h}_{\psi,b}^{(\vartheta)}\big|^2  \right\}
\end{align}
which can be further written as
\begin{align}\label{Xi1_second}
    \Xi_1 =   \sum_{\ell=1}^L   \mathbb{E} \left\{\big| {h}_{\psi,b}^{(\ell)} \big|^4 
   \right\}  +  \sum_{\ell=1}^L  \sum_{\vartheta=1, \vartheta \neq \ell}^L \mathbb{E} \left\{ \big| {h}_{\psi,b}^{(\ell)} \big|^2 
 \big|{h}_{\psi,b}^{(\vartheta)}\big|^2 
   \right\}.
\end{align}
For $ \mathbb{E} \big\{\big| {h}_{\psi,b}^{(\ell)} \big|^4 
   \big\}$ in \eqref{Xi1_second}, we have that
\begin{align}\label{Xi1_1_eq}
     \mathbb{E} &\left\{\big| {h}_{\psi,b}^{(\ell)} \big|^4 
   \right\} = {\rm Var} \left\{\big| {h}_{\psi,b}^{(\ell)} \big|^2 
   \right\} + \left( \mathbb{E} \left\{\big| {h}_{\psi,b}^{(\ell)} \big|^2
   \right\} \right)^2 \notag\\
   &\hspace{1cm} \overset{(a)}{=} \left( 4 \beta^2 + 4 \beta \Omega +\frac{\Omega^2}{m} \right) + \big( 2\beta + \Omega \big)^2,
\end{align}   
where step $(a)$ follows directly from \cite[Prop. 5.1]{EURECOM+7083}.

For $\mathbb{E} \big\{ | {h}_{\psi,b}^{(\ell)} |^2 |{h}_{\psi,b}^{(\vartheta)}|^2  \big\}$ in \eqref{Xi1_second}, we have that
\begin{align}\label{Xi1_2_eq}
    & \mathbb{E} \left\{ \big| {h}_{\psi,b}^{(\ell)} \big|^2 
 \big|{h}_{\psi,b}^{(\vartheta)}\big|^2 
   \right\}  \notag\\
   &=  \mathbb{E} \left\{ \big| Z_{\psi,b} \exp\big(\jmath \theta_{\psi,b}^{(\ell)} \big) + {\bf h}_{\psi,b}'(\ell) \big|^2 \right.\notag\\
   & \hspace{2.5cm} \left.  \times \big|{Z}_{\psi,b} \exp\big(\jmath \theta_{\psi,b}^{(\vartheta)} \big) + {\bf h}_{\psi,b}'(\vartheta) \big|^2 
   \right\}  \notag\\
   &=\mathbb{E} \left\{ Z_{\psi,b}^4  +  Z_{\psi,b}^2 |{\bf h}_{\psi,b}'(\ell)|^2  \right.  \notag\\ 
   & \hspace{1.5cm}\left. + Z_{\psi,b}^2 |{\bf h}_{\psi,b}'(\vartheta)|^2  +|{\bf h}_{\psi,b}'(\ell)|^2   |{\bf h}_{\psi,b}'(\vartheta)|^2 \right\} \notag\\
   & = \Big(1+\frac{1}{m} \Big) \Omega^2 + 4 \beta \Omega  + 4 \beta^2,
\end{align}
where ${\bf h}_{\psi,b}' \sim \mathcal{CN}( {\bf 0}_L, 2\beta {\bf I}_L )$ denotes the NLOS  channel vector at user ${\rm U}_{\psi,b}$, and ${\bf h}_{\psi,b}'(\ell)$ denotes the $\ell$-th element.
Applying \eqref{Xi1_1_eq} and \eqref{Xi1_2_eq} to \eqref{Xi1_second} leads directly to \eqref{X11_prop_eq}.

Referring to the definition of $\Xi_2$ in \eqref{Xi_def_eq}, we first obtain
\begin{align}\label{Xi2_initial}
    &\Xi_2 = \mathbb{E} \{ |\mathbf{h}_{\psi,b}^T  \hat{\bf h}_{\psi,b'}^*|^2 \}
    =  \mathbb{E} \big\{ \hat{\bf h}_{\psi,b'}^T \mathbf{h}_{\psi,b}^*  \mathbf{h}_{\psi,b}^T  \hat{\bf h}_{\psi,b'}^* \big\} \notag\\
    & = \mathbb{E}\left\{  \left(  \sum_{\ell = 1}^L \hat{h}_{\psi,b'}^{(\ell)}   \left(h_{\psi,b}^{(\ell)}\right)^* \right)   \left( \sum_{\vartheta=1}^L h_{\psi,b}^{(\vartheta)}  \left( \hat h_{\psi,b'}^{(\vartheta)} \right)^* \right)   \right\} \notag\\
    & = \sum_{\ell = 1}^L   \mathbb{E}\left\{  \big| h_{\psi,b}^{(\ell)} \big|^2 \right\} \mathbb{E}\left\{  \big| \hat h_{\psi,b'}^{(\ell)} \big|^2 \right\} \notag\\
    & \hspace{0.5cm}+ \sum_{\ell = 1}^L \sum_{\vartheta=1, \vartheta \neq \ell}^L
   \mathbb{E}\left\{  \hat{h}_{\psi,b'}^{(\ell)}     \left( \hat h_{\psi,b'}^{(\vartheta)} \right)^*  \right\}   
    \mathbb{E}\left\{  h_{\psi,b}^{(\vartheta)}  \left(h_{\psi,b}^{(\ell)}\right)^*   \right\} .
\end{align}
For $ \mathbb{E}\big\{  h_{\psi,b}^{(\vartheta)}  \big(h_{\psi,b}^{(\ell)}\big)^*   \big\}$ in \eqref{Xi2_initial}, we have
\begin{align}\label{Xi2_medium}
    &\mathbb{E}\left\{  h_{\psi,b}^{(\vartheta)}  \left(h_{\psi,b}^{(\ell)}\right)^*   \right\}  \notag\\
   & =  \mathbb{E}\left\{ \left( Z_{\psi,b} \exp\big(\jmath \theta_{\psi,b}^{(\vartheta)} \big) + {\bf h}_{\psi,b}'(\vartheta) \right)   \right. \notag\\
    & \hspace{2cm}\times\left. \left( Z_{\psi,b} \exp\big(-\jmath \theta_{\psi,b}^{(\ell)} \big) + \left( {\bf h}_{\psi,b}'(\ell) \right)^*   \right)\right\} \notag\\
    & \overset{(a)}{=} 0,
\end{align}
where the step $(a)$ follows from the fact that $\theta_{\psi,b}^{(\ell)}$ and  $\theta_{\psi,b}^{(\vartheta)}$ i.i.d. uniformly distributed over $(0, 2 \pi]$ and ${\bf h}_{\psi,b}' \sim \mathcal{CN}( {\bf 0}_L, 2 \beta {\bf I}_L )$, and which leads to 
\begin{align}\label{neq_term}
      \mathbb{E}\left\{  \hat{h}_{\psi,b'}^{(\ell)}     \left( \hat h_{\psi,b'}^{(\vartheta)} \right)^*  \right\}   
    \mathbb{E}\left\{  h_{\psi,b}^{(\vartheta)}  \left(h_{\psi,b}^{(\ell)}\right)^*   \right\} = 0.
\end{align}

Invoking \cite[Prop.~5.1]{EURECOM+7083}, we have
\(\mathbb{E}\{|h_{\psi,b}^{(\ell)}|^{2}\}=2\beta+\Omega\) and
\(\mathbb{E}\{|\hat h_{\psi,b'}^{(\ell)}|^{2}\}=2\beta+\sigma_e^{2}+\Omega\).
Combining these with \eqref{neq_term}, we finally obtain \eqref{Xi2_prop_eq}.

%Substituting \eqref{Xi2_medium} into \eqref{Xi2_initial} yields
%\begin{align}
%    \Xi_2 =& L (2 \beta + \Omega) (2 \beta + \sigma_e^2 + \Omega)  + \left( \frac{\Gamma(m+0.5)}{\Gamma(m)} \sqrt{\frac{\Omega}{m}} \right)^4  \notag\\
%    & \times  \sum_{\ell = 1}^L \sum_{\vartheta=1, \vartheta \neq \ell}^L  \mathbb{E}\left\{ \exp\Big( \jmath \big( \theta_{\psi,b'}^{(\ell)}  -  \theta_{\psi,b}^{(\ell)} + \theta_{\psi,b}^{(\vartheta)} - \theta_{\psi,b'}^{(\vartheta)}  \big)  \Big) \right\}  \label{Xi2_medium2}
%\end{align}

%Define $A_{\ell} \triangleq \exp\big(\jmath \big(\theta_{\psi,b'}^{(\ell)}  -  \theta_{\psi,b}^{(\ell)} \big)\big)$. We can rewrite the summation term in \eqref{Xi2_medium2} as
%\begin{align}\label{sum_phase}
%   &\sum_{\ell = 1}^L \sum_{\vartheta=1, \vartheta \neq \ell}^L  \exp\Big( \jmath \big( \theta_{\psi,b'}^{(\ell)}  -  \theta_{\psi,b}^{(\ell)} + \theta_{\psi,b}^{(\vartheta)} - \theta_{\psi,b'}^{(\vartheta)}  \big)  \Big) \notag\\
%   &\hspace{0.5cm}=\sum_{\ell = 1}^L \sum_{\vartheta=1, \vartheta \neq \ell}^L  A_{\ell} A_{\vartheta}^* = \left| \sum_{\ell=1}^L A_\ell \right|^2 - \sum_{\ell=1}^L \big| A_\ell \big|^2 \notag\\
%   & \hspace{0.5cm} = \left| \sum_{\ell=1}^L \exp\Big( \jmath \big(\theta_{\psi,b'}^{(\ell)}  -  \theta_{\psi,b}^{(\ell)} \big) \Big) \right|^2 - L
%\end{align}
%By using \eqref{sum_phase} in \eqref{Xi2_medium2}, we finally derive  \eqref{Xi2_prop_eq}.

\section{Proof of Theorem~\ref{VCC_thm}}\label{VCC_thm_proof}
By considering \cite[Lem. 1]{QiZhang}, the average (effective) rate for user ${\rm U}_{\psi,b}$ can be tightly approximated as
    \begin{align}\label{rate_ave}
        &\bar R_{\psi,b}  = \xi_{G,Q} \mathbb{E}\{ \log_2 ( 1 + \text{SINR}_{\psi,k} )  \}
        \notag\\
        &\approx \xi_{G,Q} \log_2\!\Bigg( \! 1 +  \frac{ \alpha^2 \mathbb{E}\{ | \mathbf{h}_{\psi,b}^T \hat {\bf h}_{\psi,b}^*|^2 \} }{ 1 + \alpha^2  \sum\limits_{b' \in [Q] \setminus \{b\} } \mathbb{E} \{ |\mathbf{h}_{\psi,b}^T  \hat{\bf h}_{\psi,b'}^*|^2 \} } \Bigg).
    \end{align}
In \eqref{rate_ave}, it is easy to see that $\mathbb{E} \{ |\mathbf{h}_{\psi,b}^T  \hat{\bf h}_{\psi,b'}^*|^2 \} = \Xi_2$ which has been derived in Proposition~\ref{Xi_prop}.    
For the term $\mathbb{E}\{ | \mathbf{h}_{\psi,b}^T \hat {\bf h}_{\psi,b}^*|^2 \}$ in \eqref{rate_ave}, we have that
\begin{align}\label{Numerator_Rate}
    \mathbb{E}&\{ | \mathbf{h}_{\psi,b}^T \hat {\bf h}_{\psi,b}^*|^2 \}
   =  \mathbb{E}\{\big| || \mathbf{h}_{\psi,b}||^2 +  \mathbf{h}_{\psi,b}^T \tilde {\bf h}_{\psi,b}^* \big|^2 \}  \notag\\
   &\overset{(a)}{=}  \mathbb{E}\{ || \mathbf{h}_{\psi,b}||^4 \} +  \mathbb{E}\{ \tilde {\bf h}_{\psi,b}^T  \mathbf{h}_{\psi,b}^*  \mathbf{h}_{\psi,b}^T \tilde {\bf h}_{\psi,b}^* \} \notag\\
    &= \Xi_1 + \mathbb{E}\{ {\rm Tr}\{ \mathbf{h}_{\psi,b}^*  \mathbf{h}_{\psi,b}^T \tilde {\bf h}_{\psi,b}^* \tilde {\bf h}_{\psi,b}^T  \} \} \notag\\
    &\overset{(b)}{=} \Xi_1 + {\rm Tr} \{ \mathbb{E} \{  \mathbf{h}_{\psi,b}^*  \mathbf{h}_{\psi,b}^T \} 
 \mathbb{E} \{  \tilde {\bf h}_{\psi,b}^* \tilde {\bf h}_{\psi,b}^T  \} \} \notag\\
 & = \Xi_1 + \sigma_e^2 \mathbb{E}\{ || {\bf h}_{\psi,b} ||^2 \} 
 \overset{(c)}{=}  \Xi_1 + \sigma_e^2 L (2\beta + \Omega),
\end{align}
where $(a)$ follows from the fact that $\mathbb{E}\{ ||{\bf h}_{\psi,b}||^2 {\bf h}_{\psi,b} \tilde {\bf h}_{\psi,b}^* \} =\mathbb{E}\{ ||{\bf h}_{\psi,b}||^2 {\bf h}_{\psi,b} \} \mathbb{E}\{ \tilde {\bf h}_{\psi,b}^* \} =0 $ due to $\tilde {\bf h}_{\psi,b} \sim \mathcal{CN}( {\bf 0}_L, \sigma_e^2 {\bf I}_L )$, $(b)$ follows from the interchange between the expectation and trace operators, and $(c)$ follows from the fact that $ \mathbb{E}\{ || {\bf h}_{\psi,b} ||^2 \} = L (2\beta + \Omega)$ (cf. \cite[Prop. 5.1]{EURECOM+7083}). Substituting \eqref{Numerator_Rate} into \eqref{rate_ave}, we can derive the approximation for $\bar R_{\psi,b}$.
Considering that $\bar R_{\rm sum}= \sum_{\psi \in \Psi} \sum_{b \in [Q]} \bar R_{\psi,k}$, we finally have \eqref{sumRate_ave}. Considering the effective gain defined in \eqref{Gain_def} and using the approximation for $\bar R_{\rm sum}$, we can easily derive \eqref{gain_thm}.

%%%%%%%%%%%%%%%%%%%%%%%%%%%%%%%%%%%%%%%%%%%%%%%%%%%%%%%%%%%
	\bibliographystyle{IEEEtran}				%%%%%%%%%%%%%%%%%%%%%
	\bibliography{IEEEabrv,aBiblio}			%%%%%%%%%%%%%%%%%%%%%
%%%%%%%%%%%%%%%%%%%%%%%%%%%%%%%%%%%%%%%%%%%%%%%%%%%%%%%%%%%

\end{document}